\documentclass[aps,prd,showpacs,onecolumn,nofootinbib,superscriptaddress,amsmath,amssymb,8pt]{revtex4-2}
\usepackage{amsmath,amssymb,amsfonts}
\usepackage{bm}
\usepackage{hyperref}
\usepackage{slashed,braket}
\usepackage{color}
\usepackage{graphicx,graphics}
\usepackage[title,titletoc]{appendix}
\usepackage{stackengine}

\usepackage{tikz}\usetikzlibrary{shapes.geometric, arrows}\tikzstyle{method} = [rectangle, rounded corners, minimum width=3cm, minimum height=1cm, text centered, draw=black]\tikzstyle{obj} = [rectangle, minimum height=1cm, text centered, draw=black]\tikzstyle{arrow} = [thick,->,>=stealth]

\newcommand{\cD}{\mathcal{D}}

\newcommand{\tr}{\operatorname{tr}}

\newcommand{\D}{\Delta}
\newcommand{\bp}{\mathbf{p}}

\newcommand{\be}{\begin{equation}}
\newcommand{\ee}{\end{equation}}

\newcommand{\mz}[1]{{#1}}

{\color{blue}}%
{}

\begin{document}

\title{Effective lagrangian for the macroscopic motion of fermionic matter}

\author{Maik Selch}
\email{maik.selch@t-online.de}
\affiliation{Physics Department, Ariel University, Ariel 40700, Israel}

\author{Ruslan A.~Abramchuk}
\email{abramchuk@phystech.edu}
\affiliation{Physics Department, Ariel University, Ariel 40700, Israel}
\affiliation{On leave of absence from Kurchatov Complex for Theoretical and Experimental Physics, B. Cheremushkinskaya 25, Moscow, 117259, Russia}

\author{M.A.Zubkov}
\email{mikhailzu@ariel.ac.il}
\affiliation{Physics Department, Ariel University, Ariel 40700, Israel}

\date{\today}

\begin{abstract}
    We consider macroscopic motion of quantum field systems. Zubarev statistical operator allows to describe several types of motion of such systems in thermal equilibrium. We formulate the corresponding effective theory on the language of functional integral. The effective lagrangian is calculated explicitly for the fermionic systems interacting with dynamical gauge fields. Possible applications to physics of quark - gluon plasma are discussed. 
\end{abstract}

\maketitle

\tableofcontents

\section{Introduction}

Continuous medium approximation is widely applied for modeling of  strongly-interacting relativistic systems. In high energy physics, in particular, this approach is used for the investigation  strongly-interacting Quark Gluon Plasma ((s)QGP) in Heavy Ion Collisions (HIC)
\cite{Bjorken1983,Heinz2013th,Csernai2013,Teaney2001av,Zinchenko2022tyg,Bravina2021arj,Rybalka2018uzh}. In astrophysics it is used for description of 
 neutron stars and binary systems (with general-relativistic magneto-hydrodynamic simulations) 
\cite{Neilsen2014,Galeazzi2013,Lioutas2022ghb}.
This approximation implies quasi-equilibrium of medium, i.e.  sufficiently small space-time cells of matter (`grains') are assumed to be in thermodynamic equilibrium.

Parameters for the realistic hydrodynamic models are usually derived from
the underlying matter equation of state (EoS) or directly from the quantum field theory  (QFT) 
\cite{Kharzeev2007wb,Karsch2007jc,Hempel2009mc,Shovkovy2010xk,Neilsen2014,Tuchin2011jw,Astrakhantsev2018oue,Astrakhantsev2017nrs,Buzzegoli2017cqy,Lublinsky2007mm}.
The explicit dependence of EoS on  \((T\text{ or }\epsilon, \mu\text{ or }n)\) is usually considered,
where \(T\) -- temperature, \(\epsilon\) -- (kinetic) energy density,
\(\mu_i\) -- chemical potentials and \(n_i\) -- the associated charge densities. 

Inertial forces on the scale of a grain 
\footnote{Say, the forces caused by rotation acting at the nuclear scale of a few fm} 
can be strong enough to qualitatively change equation of state 
and modify phase diagram.
Such effects naturally arise in general relativity. In particular, notions of  Tolman-Ehrenfest and Unruh temperatures \cite{Bermond2022mjo} appear this way.

For the idealized rigidly rotating hot QCD such effects were recently discovered using Lattice simulations \cite{Chernodub2022veq,Braguta2023yjn,Braguta2023kwl,Astrakhantsev2021fua,Braguta2021jgn,Braguta2020biu}. For the accelerated QFT systems these effects were discussed in \cite{Prokhorov2023dfg,Khakimov2023emy}
using the Zubarev approach \cite{Prokhorov2019cik}.
Inside neutron stars velocities reach a quarter of speed of light. Therefore, relativistic corrections to the EOS of   nuclear matter are considerable \cite{Akmal1998cf}.

Effects related to rotation in HIC recently attracted attention
because of the large values of vorticity (angular velocity) of (s)QGP 
of the order of \(\sim 10^{22}\text{ s}^{-1}\sim 10\) MeV \cite{STAR2017ckg,Csernai2013,Deng2016gyh}.
Effects of acceleration are to be investigated as well --- the values of matter acceleration in HIC are expected to be of the order of hundreds MeV \cite{Kharzeev2007wb,Labun2010wf}. These effects, however, are much less known compared to effects of vorticity \cite{Csernai2013,Baznat2017jfj,Bravina2021arj}. 

Other relativistic effects can be relevant as well for the neutron matter EoS
for modeling of a neutron star evolution. They can even be more important for  modeling binaries, e.g.~neutron star mergers,
in which the astronomically large objects coalesce in a few milliseconds \cite{Lioutas2022ghb,Kastaun2016yaf}. 
We expect that within these objects the hydrodynamic velocities, angular velocities, accelerations can influence the nuclear matter EoS.

In this paper we systematically address the problem for the case of inertial forces 
that emerge due to the macroscopic motion of the strongly interacting matter.
To perform a rigorous QFT analysis we rely on Zubarev statistical operator \cite{Zubarev1979}.
The macroscopic motion and temperature distribution are encoded in the temperature four-vector (the frigidity vector) and local chemical potentials.
Though, as a particular application we mostly keep in mind analysis of the sQGP emerging in HIC, 
our findings may be relevant for modeling of evolution of neutron stars and binaries.
Consideration of external electromagnetic fields and the chiral density is also possible within the Zubarev approach, 
and might be significant \cite{Buividovich2015jfa},  
but we leave it for later publications. 

In general, Zubarev operator allows to analyze the non-equilibrium systems.
The thermodynamic equilibrium imposes \cite{Zubarev1979,Buzzegoli2017cqy,Becattini2020qol,Buzzegoli2020ycf} restrictive conditions on the macroscopic motion and temperature distribution.
Aside from the motion with constant velocity, the equilibrium frigidity is parametrized by a constant thermal vorticity, 
which is an anti-symmetric tensor.
The axial part of thermal vorticity  
is proportional to the angular velocity of  rigid rotation of the system.  
The polar part
is proportional to the linear acceleration of the system at the initial moment in time.

In simulation of HIC, which results in description  of the (s)QGP evolution, 
the frigidity four vector can be easily extracted (see e.g. \cite{Bravina2021arj}) for each cell, or a coarse grain,
and approximated by an equilibrium frigidity value,
which comprises motion with constant velocity, rotation and accelerated motion. 

To analyze the thermodynamic state of such a grain,
we derive the effective action for the macroscopically moving grain,
which allows to compute, e.g.~with the lattice simulations, the path integral, obtain the corresponding partition function and local thermodynamic quantities.
Our effective action contains the terms for general macroscopic motion of the substance,
which have not been considered in literature before.

It is instructive to discuss separately the particular case of rigid rotation. In order to perform lattice Monte-Carlo simulations the system is typically considered in  rotating reference frame \cite{Yamamoto2013zwa,Braguta2020biu,Braguta2021jgn,Chernodub2022veq,Braguta2023kwl,Braguta2023yjn,Landsteiner2020xuv}. In these simulations in order to avoid the sign problem, the angular velocity is taken to be imaginary, at the end of calculations the analytical continuation to real values is to be performed. 
In this approach, which is called here ``passive'' description of rotation, 
the vacuum is rotating by definition.
Notice that the non-trivial metrics are introduced for many purposes while considering macroscopic motion of substance \cite{Chernodub2021nff}.

Another approach, which is called here ``active'' description of rotation,
is to define the statistical operator of a rotating system,
which was done rigorously in \cite{Zubarev1979} for any kind of macroscopic motion that is admitted for thermal equilibrium. In 
 \cite{VilenkinCVE} the rigid rotation has been considered within the similar approach. This approach has been developed later e.g. in \cite{Landsteiner2018era,Landsteiner2020xuv}.
For the "active" description of rotation  vacuum is defined in an inertial frame, while excitations above vacuum are rotated (either thermal or those caused by chemical potential).
Notice that duality of the "active" and "passive" descriptions emerges also in non-relativistic physics \cite{Stone1999gi}.

It appears that the rotating vacuum is identical to the static vacuum in an IR-complete theory of free Dirac fermions,
as was shown, for example, in \cite{Abramchuk2018jhd}.
In the present paper we demonstrate, in particular (see Appendix \ref{comparison}), 
that the effective action (for the gauge field interacting with fermions) obtained from the active description of rotation
coincides with the action obtained from the passive description of rotation \cite{Yamamoto2013zwa},
which is not obvious \emph{a priori}. 

Another type of motion admitted for thermal equilibrium according to the analysis of Zubarev \cite{Zubarev1979} is the motion that comprises linear acceleration at the initial moment (see Appendix \ref{accelerate}). Notice that this is not motion with constant acceleration that does not depend on time, its description in more complicated at later stages. This kind of motion has been investigated analytically (see, for example, \cite{Prokhorov2023dfg,Khakimov2023emy}). However, to the best of our knowledge, equilibrium QCD with such kind of macroscopoc motion was not investigated using lattice simulations.
Interacting theories with Rindler metric (accelerating systems) were considered only recently \cite{Akhmedov2021agm}.
In the present paper we propose, in particular, the effective action for the equilibrium system that admits linear acceleration. It appears that the effective lagrangian for this type of motion coincides with the lagrangian of the system at rest. However, the corresponding  temperature becomes depending on the spatial coordinates. 
The most general case of equilibrium motion contains superposition of motion with constant linear velocity, rigid rotation  and accelerated motion. We derive the effective action \eqref{finL}, which may be used for description of such a motion for any system of fermions interacting with gauge field, including QCD. In particular, we expect that this effective action may be useful for lattice QCD simulations.

\section{Zubarev statistical operator and motion of substance in equilibrium}
\label{SectZubarev}

In this paper we assume metric signature $(+,-,-,-)$.
Following \cite{Zubarev1979} we present the covariant form for the statistical operator which provides candidates for a proper description of macroscopic motion of a substance in global thermodynamical equilibrium. Namely, the logarithm of the statistical operator is expressed as

\begin{align}
\log\hat \rho =-\alpha -\mz{\int d\Sigma \beta n_{\nu}(\hat{T}^{\nu\rho}u_{\rho}-\sum_i\mu_i\hat{j}_i^{\nu})}.\label{Zub}
\end{align}
Here integration is over three - dimensional space - like hypersurface $\Sigma$, while $\hat{T}^{\nu\rho}$ is the gravitational (or, Belinfante - Rosenfeld) stress - energy tensor operator. By $d \Sigma$ we denote element of integration. The function $\beta(x)$ may depend on coordinates as well as $n(x)$ and $u(x)$.  The latter vectors obey $$
n^2(x)=u^2(x)=1
$$
Four - vector $n(x)$ is orthogonal to surface $\Sigma$, while $u$ may be interpreted as the macroscopic four - velocity. Function $\beta(x)$ may be interpreted as inverse temperature depending on coordinates.
The combination \(\beta_\mu(x)=\beta(x)u_\mu(x)\) is the frigidity vector.

Space - time considered here is flat Minkowski space, which admits foliation into the three - dimensional space - like hypersurfaces $\Sigma_\sigma$ depending on parameter $\sigma$. We consider evolution of the system in the parameter $\sigma$ with its initial value $\sigma_i$ and final value $\sigma_f$.

{The boundary conditions for the evolution are the total translational, angular and boost momentum and charges 
\[P^\mu_{i,f} = \braket{\left.\int d\Sigma_\nu\hat T^{\mu\nu}\right|_{\sigma_{i,f}}},\]\[ M^{\mu\nu}_{i,f} = \braket{\left.\int d\Sigma_\rho (x^{\mu}\hat T^{\rho\nu}-x^{\nu}\hat T^{\rho\mu})\right|_{\sigma_{i,f}}},\]\[
  ~ Q_{i,f}^k = \braket{\left.\int d\Sigma_\nu\hat j^{\nu}_k\right|_{\sigma_{i,f}}}.\]
The gravitational (or, Belinfante-Rosenfeld) energy momentum tensor in the Zubarev statistical operator comprises all ten Poincar\'e charges, the four translational charges with canonical energy momentum tensor $\hat T^{\mu\nu}_{can}$ as Noether current and Lorentz transformation charges with canonical Lorentz transformation tensor 
\begin{align}
\hat M^{\mu\nu\lambda}_{can}=(x^{\nu} \hat T^{\mu\lambda}_{can\,}-x^{\lambda}\hat T^{\mu \nu}_{can\,})+\hat S^{\mu \nu \lambda}_{\,\,\,\,}.
\end{align}
The former term comprises angular and boost momentum contributions, while the latter term is the spin current. It is related to the anti-symmetric part of the canonical energy momentum tensor by
\begin{align}
D_{\mu}\hat S^{\mu}_{\,\,\,\,\nu\lambda}=\hat T^{can}_{\lambda\nu}-\hat T^{can}_{\nu\lambda}.
\end{align}
with covariant derivative $D_{\mu}$. The (symmetric) gravitational (or, Belinfante-Rosenfeld) energy momentum tensor may then be expressed in terms of the canonical energy momentum tensor and the spin current by
\begin{align}
\hat T^{\mu\nu}=\hat T_{can}^{\mu\nu}+\frac{1}{2}D_{\lambda}(\hat S^{\mu\nu\lambda}+\hat S^{\nu\mu\lambda}-\hat S^{\lambda\nu\mu}).
\end{align}}

The stationarity condition \(\frac{d\hat\rho}{d\sigma} =0\)
includes the thermodynamic equilibrium,
and requires the integrand to be divergence-free \cite{Zubarev1979}.
In order to have equilibrium we should require that expression of Eq. (\ref{Zub}) does not depend on $\Sigma = \Sigma_s$. 
For this it is sufficient if the right - hand side of Eq. (\ref{Zub}) does not depend on the form of $\Sigma$ at all. 
This requirement is equivalent to 
\begin{align}
	0&=\partial_{\nu}\beta(\hat T^{\nu\rho}u_{\rho}-\sum_i\mu_i\hat j_i^{\nu})\nonumber\\ &=\hat T^{\nu\rho}\partial_{\nu}\beta u_{\rho}-\sum_i\hat j_i^{\nu}\partial_\nu\beta\mu_i.\label{Zub2}
\end{align}
This equation is satisfied by 
\begin{align}
	\beta\mu_i&=\zeta_i={\rm const} \nonumber\\  \beta_\rho = \beta u_{\rho} &= b_\rho + \bar\omega_{\rho\sigma} x^\sigma \label{Zub3}
\end{align}
with constant anti - symmetric tensor $\bar\omega_{\rho\sigma}$ (the thermal vorticity). 
{Then
\begin{equation}
\beta(x) = \sqrt{b^2 + g^{\mu\rho}\bar\omega_{\mu\nu}\bar\omega_{\rho\sigma} x^\nu x^\sigma + 2 g^{\mu\rho}b_\mu  \bar\omega_{\rho\sigma} x^\sigma}
\end{equation}
and
\begin{equation}
u_\mu(x) = \frac{b_\rho + \bar\omega_{\rho\sigma} x^\sigma}{\sqrt{b^2 + g^{\mu\rho}\bar\omega_{\mu\nu}\bar\omega_{\rho\sigma} x^\nu x^\sigma + 2 g^{\mu\rho}b_\mu  \bar\omega_{\rho\sigma} x^\sigma}}
\end{equation}
One can define the four - "acceleration" vector 
\begin{equation}
a_\mu =\frac{1}{\beta}\bar\omega_{\nu\mu}u^\nu= \frac{\bar\omega_{\nu\mu}(b^\nu + \bar\omega^\nu_{\,.\,\sigma} x^\sigma)}{b^2 + g^{\mu\rho}\bar\omega_{\mu\nu}\bar\omega_{\rho\sigma} x^\nu x^\sigma + 2 g^{\mu\rho}b_\mu  \bar\omega_{\rho\sigma} x^\sigma} \label{Zub3}
\end{equation}
and vorticity
\begin{equation}
\omega_\mu = -\frac{1}{2\beta}\epsilon_{\mu\nu\rho\sigma}u^\nu \bar\omega^{\rho\sigma}$$$$=-\frac{\epsilon_{\mu\nu\rho\sigma}(b^\nu + \bar\omega^\nu_{\,.\,\tau} x^\tau) \bar\omega^{\rho\sigma}}{2({b^2 + g^{\mu\rho}\bar\omega_{\mu\nu}\bar\omega_{\rho\sigma} x^\nu x^\sigma + 2 g^{\mu\rho}b_\mu  \bar\omega_{\rho\sigma} x^\sigma})}
\end{equation}
Chemical potential receives the form
\begin{equation}
\mu_i(x)=\frac{\zeta_i}{\beta(x)} = \frac{\zeta_i}{\sqrt{b^2 + g^{\mu\rho}\bar\omega_{\mu\nu}\bar\omega_{\rho\sigma} x^\nu x^\sigma + 2 g^{\mu\rho}b_\mu  \bar\omega_{\rho\sigma} x^\sigma}}
\end{equation}}
The limiting case of non - relativistic macroscopic motion corresponds to $|b_\rho| \gg |\bar\omega_{\rho\sigma} x^\sigma|$. In this case we have approximate relation:
$$
\beta(x) \approx \| b\| = {\rm const}
$$
In the reference frame, in which spatial components of $b$ vanish, we obtain:
$$
b \approx \beta (1,0,0,0)
$$
and
$$
\bar{\omega}_{\mu\nu} =\beta ( \epsilon_{\mu\nu\alpha 0}\omega^\alpha  + a_\mu \delta^0_\nu - a_\nu \delta^0_\mu)
$$
while
$$
\beta\mu_i=\zeta_i
$$

In the general case the tensor $\bar{\omega}$  may be decomposed as 
\begin{equation}
\bar{\omega}_{\mu\nu} =\beta ( \epsilon_{\mu\nu\alpha\beta}\omega^\alpha u^\beta + a_\mu u_\nu - a_\nu u_\mu)
\end{equation}
The following integrals of motion enter the expression for the statistical operator:
\begin{eqnarray}
	\hat P^\mu &=& \int d\Sigma n_\nu \hat T^{\nu \mu}\nonumber\\
	\hat Q_i & = & \int d\Sigma n_\nu \hat j_i^\nu\nonumber\\
	\hat J^{\nu\mu} &=& \int d\Sigma n_\rho (\hat T^{\rho \mu} x^\nu - \hat T^{\rho \nu} x^\mu ) 
\end{eqnarray}
We obtain:
\begin{equation}
	\rho = \frac{1}{Z}e^{-b_\mu \hat P^\mu + \frac{1}{2}\bar{\omega}_{\mu\nu} \hat J^{\mu\nu} + \sum_i \zeta_i \hat Q_i}\label{rho00}
\end{equation}
Tensor $J^{\mu\nu}$ may be decomposed as
\begin{equation}
	\hat J^{\mu\nu} = \epsilon_{\mu\nu\alpha\beta}\hat J^\alpha u^\beta - \hat K_\mu u_\nu + \hat K_\nu u_\mu 
\end{equation}
Here $K_\mu$ is the generator of boost while $J_\nu$ is generator of rotation (both are taken in the co - moving reference frame). In terms of these generators we obtain the following expression for the statistical operator:
 \begin{equation}
 \hat	\rho = \frac{1}{Z}e^{-b_\mu \hat P^\mu - \beta a_\mu \hat K^\mu  + \beta {\omega}_{\mu} \hat J^{\mu} + \sum_i \zeta_i \hat Q_i}\label{rho0}
 \end{equation}
Alternatively one can reduce tensor $\hat J_{\mu\nu}$ to momentum $\hat J^{(0)}$ and boost generator $\hat K^{(0)}$ in laboratory reference frame: 
\begin{equation}
	\hat J^{\mu\nu} = \epsilon_{\mu\nu\alpha\beta}\hat J^{(0)\alpha} n^\beta - \hat K^{(0)}_\mu n_\nu + \hat K^{(0)}_\nu n_\mu 
\end{equation} 
Let us consider the particular case, when hypersurface $\Sigma$ is the hyperplane $t = 0$ in the inertial reference frame. Then $n = (1,0,0,0)$. Besides, we assume $\bar{\omega}_{0i} = 0$ and $b = (\beta_0,0,0,0)$. At the same time the rigid rotation is present given by angular velocity $\vec{\omega}$ (see Appendix \ref{rigid}) \footnote{Notice that in this situation acceleration of Eq. (\ref{Zub3})  does not vanish in spite of naive expectations. Namely, we have $a^\mu = (0,\vec{a})$ with 
$$
\vec{a} = \frac{\omega^2 \vec{r}_\bot}{1 - \omega^2 r^2_\bot}
$$}.
 Eq. (\ref{rho00}) is reduced in this case to 
 \begin{equation}
\hat	\rho = \frac{1}{Z}e^{- \beta_0 \hat P^0   + \beta_0 {\omega}^{(0)}_{\mu} \hat J^{\mu} + \sum_i \mu^{(0)}_i \hat Q_i}\label{rho001}
\end{equation}
with $\mu^{(0)}_i = \zeta_i/\beta_0$ and
$\omega^{(0)}_\mu = (0, \vec{\omega})$.   
In this case rotation appears as an enhancement of angular momentum, and results in appearance of the product of angular velocity and angular momentum projected to the axis of rotation.

In the following in this paper we do not restrict ourselves by this particular case. {\it But we consider only the case, when surface $\Sigma$ entering Zubarev statistical operator is the hyperplane of constant time in the inertial Laboratory reference frame, which implies $n^{\mu}=(1,0,0,0)^T$.}   

\section{Non - interacting Dirac fermions}
\subsection{Zubarev statistical operator for non - interacting Dirac fermions}

We proceed to calculate the integrands in statistical operator explicitly for the case of massive Dirac fermions. We consider fermion vector current operator $j^{\rho}(x)=\overline{\Psi}(x)\gamma^{\rho}\Psi (x)$ with chemical potential $\mu$. 

The Dirac Lagrangian for massive classical complex - valued fermionic fields $\Psi$ is given by

\begin{align}
	\mathcal{L}(\overline{\Psi}(x),\Psi (x))&=\overline{\Psi}(x)(i\gamma^{\mu}\partial_{\mu}-m)\Psi (x)\\
	&=\overline{\Psi}(x)(\frac{i}{2}\gamma^{\mu}\overset{\leftrightarrow}{\partial}_{\mu}-m)\Psi (x)
\end{align}

with Lorentzian $\gamma$-matrices $\{ \gamma^{\mu},\gamma^{\nu}\}=2g^{\mu\nu}$ and metric signature $(+,-,-,-)$. Here 
$$
\overset{\leftrightarrow}{\partial}_{\mu} = \overset{\rightarrow}{\partial}_{\mu} - \overset{\leftarrow}{\partial}_{\mu}
$$
 The two forms are equivalent in the absence of boundary terms (which we assume here). We will use the former version to derive the canonical commutation relations as well as the Dirac Hamiltonian (density) but the latter (more symmetric form) in the course of the calculations. The canonically conjugate momenta are given by 

\begin{align}
	\Pi_{\Psi}=\frac{\partial\mathcal{L}}{\partial (\partial_0\Psi)}=i\overline{\Psi}\gamma^0,\,\,\,\, \Pi_{\overline{\Psi}}=\frac{\partial\mathcal{L}}{\partial (\partial_0\overline{\Psi})}=0.
\end{align}

This implies the canonical (anti) commutation relations for the corresponding operators in quantum theory 

\begin{align}
	\{(\hat \Pi_{\Psi})_{\alpha}(x),\hat \Psi_{\beta}(y)\}=i\delta^3(x-y)\delta_{\alpha\beta} \Leftrightarrow \nonumber\\ \{\hat{\Psi}_{\alpha}(x),\hat{\overline{\Psi}}_{\beta}(y)\}=\delta^3(x-y)\gamma^0_{\alpha\beta}.
\end{align}

The Hamiltonian is given by $\hat H=\int d\Sigma \hat{\mathcal{H}}$.
The Hamiltonian density is given by

\begin{align}
	\hat{\mathcal{H}}&=\hat{\Pi}_{\Psi}(\partial_0\hat{\Psi})+(\partial_0\overline{\hat{\Psi}})\hat{\Pi}_{\overline{\hat{\Psi}}}-\mathcal{L}(\hat{\overline{\Psi}},\hat{\Psi} )\nonumber\\&=-\hat{\overline{\Psi}}(\frac{i}{2}\gamma^j\overset{\leftrightarrow}{\partial}_j)\hat{\Psi} + m \hat{\overline{\Psi}}\hat{\Psi} \label{calH}
\end{align}

In the quantum theory we substitute complex - valued field $\Psi$ by operator - valued field $\hat{\Psi}$;   operators $\hat O$ satisfy the Heisenberg equation of motion

\begin{align}
	i\partial_0\hat{O}=-[\hat{H},\hat{O}],
\end{align}
where $\hat H$ is the Hamiltonian operator defined above. In Eq. (\ref{calH}) we imply normal ordering. The commutation relations we will need subsequently are those for $\hat{O}=\hat\Psi_{\alpha} ,{\overline{\hat\Psi}}_{\alpha}$. They take the form
\begin{align}
	&[\hat H,\overline{{\hat{\Psi}}}_{\alpha}(x)]=\mz{}\overline{{\hat{\Psi}}}_{\beta}(x)(i\gamma^j_{\beta\gamma}\overset{\leftarrow}{\partial}_j+m\delta_{\beta\gamma})\gamma^0_{\gamma\alpha}\\
	&[\hat H,{\hat{\Psi}}_{\alpha}(x)]=\mz{}\gamma^0_{\alpha\beta}(i\gamma^j_{\beta\gamma}\overset{\rightarrow}{\partial}_j-m\delta_{\beta\gamma}){\hat{\Psi}}_{\gamma}(x).
\end{align}
In order to make the calculations of the density operators explicit we need the symmetrized canonical energy momentum tensor operator which is given by
\begin{align}
\hat	T^{can}_{\mu\nu}=\overline{{\hat{\Psi}}}\frac{i}{2}\gamma_{(\mu}\overset{\leftrightarrow}{\partial}_{\nu )}{\hat{\Psi}} -g_{\mu\nu}\mathcal{L}(\overline{{\hat{\Psi}}},{\hat{\Psi}} ).
\end{align}
Here 
$$
\gamma_{(\mu}\overset{\leftrightarrow}{\partial}_{\nu )} = \frac{1}{2}(\gamma_{\mu}\overset{\leftrightarrow}{\partial}_{\nu } + \overset{\leftrightarrow}{\partial}_{\nu }\gamma_{\mu}) = 
$$
$$
= \frac{1}{2}(\gamma_{\mu}\overset{\rightarrow}{\partial}_{\nu } + \overset{\rightarrow}{\partial}_{\nu }\gamma_{\mu}) - \frac{1}{2}(\gamma_{\mu}\overset{\leftarrow}{\partial}_{\nu } + \overset{\leftarrow}{\partial}_{\nu }\gamma_{\mu}) 
$$ 
{The anti-symmetric part is connected to the spin tensor for Dirac fermions which reads
\begin{align}
\hat S^{\mu}_{\, .\, \nu\lambda}=\frac{i}{8}\hat{\Psi}\{\gamma^{\mu},[\gamma_{\nu},\gamma_{\lambda}]\}\Psi .
\end{align}
}On classical equations of motion for ${\hat{\Psi}}$ we have $\mathcal{L}(\overline{{\hat{\Psi}}},{\hat{\Psi}} )=0$. This results in
	\begin{align}
	\hat	T^{can}_{\mu\nu}=\overline{{\hat{\Psi}}}\frac{i}{2}\gamma_{(\mu}\overset{\leftrightarrow}{\partial}_{\nu )}{\hat{\Psi}} .
	\end{align}
	The same expression is obtained if we define stress - energy tensor through variation of action with respect to metric tensor.
In the following calculations the derivatives with overset arrows are always meant to act either on ${\hat{\Psi}}$ or on $\overline{{\hat{\Psi}}}$.  The integrands of the ansatz for the density operator may be rewritten as follows
	\begin{align}
		\nonumber &\int \beta d\Sigma n^{\nu}g_{\nu\rho}(\hat T_{can}^{(\rho\sigma )}u_{\sigma}-\sum_i\mu_i \hat j_i^{\rho})\\
		\nonumber &=\int \beta d\Sigma n^{\rho}
		((\overline{{\hat{\Psi}}}\frac{i}{2}\gamma_{(\rho}\overset{\leftrightarrow}{\partial}_{\sigma )}{\hat{\Psi}} )u^{\sigma}-\sum_i\mu_i \hat j_i^{\rho})\end{align}
	We arrive at (space derivatives are denoted by Latin letters):
	\begin{align}
	&	{\cal R}[\beta(x),\mu_i(x),u^\mu(x),\hat\Psi(x),\bar{\hat\Psi}(x)] \equiv 	-{\rm ln}\, \hat \rho -\alpha \nonumber\\ &  =\int \beta d\Sigma (\overline{{\hat{\Psi}}}\frac{i}{4}\gamma_{0}\overset{\leftrightarrow}{\partial}_{j}{\hat{\Psi}} u^{j} -\sum_i\mu_i \hat j_i^{0}\nonumber\\
		 &+u^{j}([\hat H,\overline{{\hat{\Psi}}}]\frac{1}{4}\gamma_{j}{\hat{\Psi}}-\overline{{\hat{\Psi}}}\frac{1}{4}\gamma_{j}[\hat H,{\hat{\Psi}} ]))\nonumber\\ &+u^{0}([\hat H,\overline{{\hat{\Psi}}}]\frac{1}{2}\gamma_{0}{\hat{\Psi}}-\overline{{\hat{\Psi}}}\frac{1}{2}\gamma_{0}[\hat H,{\hat{\Psi}} ]))\nonumber\\
		\nonumber &=\int \beta d\Sigma (\overline{{\hat{\Psi}}}\frac{i}{4}\gamma_{0}\overset{\leftrightarrow}{\partial}_{j}{\hat{\Psi}} u^{j}  -\sum_i\mu_i \hat j_i^{0}+u_{k}(\overline{{\hat{\Psi}}}(i\gamma^j\overset{\leftarrow}{\partial}_j+m)\gamma^0\frac{1}{4}\gamma^k{\hat{\Psi}}\\
		\nonumber &+\overline{{\hat{\Psi}}}\frac{1}{4}\gamma^{k}(-\gamma^0(i\gamma^j\overset{\rightarrow}{\partial}_j-m){\hat{\Psi}} ))+u_{0}(\overline{{\hat{\Psi}}}(i\gamma^j\overset{\leftarrow}{\partial}_j+m)\gamma^0\frac{1}{2}\gamma^{0}{\hat{\Psi}}\\
		\nonumber &+\overline{{\hat{\Psi}}}\frac{1}{2}\gamma^0(-\gamma^0(i\gamma^j\overset{\rightarrow}{\partial}_j-m){\hat{\Psi}} ))).
	\end{align}
	Notice that here and below the derivatives act on the fields ${\hat{\Psi}}$ and ${{\hat{\Psi}}}^\dagger$ only, and do not act on $n$ and $u$. Besides, $\bar{{\hat{\Psi}}}$ is defined as $\bar{{\hat{\Psi}}} = {\hat{\Psi}}^\dagger \gamma^0$.

\subsection{Representation of partition function in the form of functional integral}

We can introduce the notion of coherent state associated with the Grassmann - valued field $\psi$ defined along the surface $\Sigma$:
\begin{equation}
|\phi\rangle = e^{\int d\Sigma n_\mu \hat{\Psi}^\dag \gamma^0 \gamma^\mu \phi} |\Omega\rangle , \quad \langle \tilde\phi | =\bra{\Omega} e^{\int d\Sigma n_\mu \tilde{\phi} \gamma^0 \gamma^\mu \hat \Psi} 
\end{equation}
Here $\phi(x)$ and $\tilde{\phi}(x)$ are the different Grassmann - valued fields.  In the following we will also use notation $\bar{\phi} = \tilde{\phi} \gamma^0$. For brevity below in the text of the present paper we assume $\bra \phi = \bra{\tilde{\phi}}$. However, notice that formally, bra and ket vectors  $\bra{\tilde{\phi}}$ and $\ket{\phi}$ are not conjugate to each other. (These vectors would be conjugate to each other if $\tilde{\phi}$ is considered as conjugate to $\phi$.) $|\Omega \rangle$ is the Fock space state, in which all one - particle states are vacant, while all anti - particle states are occupied. Notice that this is not the conventional vacuum, but it is more useful for our purposes to construct the tower of coherent states above this state rather than above the conventional vacuum.   
These coherent states obey the following properties:
\begin{enumerate}
	\item 
	
	$$
	\hat\Psi(x) |\psi\rangle = \hat\Psi(x) e^{\int d\Sigma n_\mu \hat \Psi^+ \gamma^0 \gamma^\mu \psi} |\Omega\rangle = \psi(x) |\psi\rangle$$
	
	\item
	
	$$
	\langle \phi |\psi\rangle = e^{\int d\Sigma n_\mu \bar\phi  \gamma^\mu \psi} 
	$$
	
	\item
	
	\begin{equation}
	1 = \int D\bar\psi D\psi e^{-\int d\Sigma n_\mu \bar{\psi} \gamma^\mu \psi} |\psi\rangle\langle \psi |\label{1}
	\end{equation}
	
\end{enumerate}
These expressions are proven in Appendix \ref{SectAppCoherent}. 

We fix surface $\Sigma$ as the hypersurface $t = 0$ in Minkowski space - time and represent the Zubarev statistical operator defined on $\Sigma$ as 
$$
\hat \rho = e^{-\alpha} \lim_{N\to \infty}\Pi_{s=0,1,\ldots,N-1}e^{-{\cal R}[\beta(x),\mu_i(x),u^\mu(x),\hat\Psi(x),\bar{\hat\Psi}(x)]\delta(N)},$$$$ \, \delta(N)=1/N 
$$

 Next, we insert unity from Eq. (\ref{1}) between each two multipliers in the above product, and arrive at the expression for  the partition function 
\begin{eqnarray}
	&& Z[n(x),u(x),\beta(x),\mu_i(x)]= e^{\alpha} = \nonumber\\&&
	= \int D\bar{\phi} D\phi \, e^{\int_0^1 d \tau \int d\Sigma  {L}(\overline{\phi},\phi )}
\end{eqnarray}
 Now $\phi(x,\tau)$ and $\bar{\phi}(x)=\tilde{\phi}(x)\gamma^0$ are independent Grassmann - valued fields depending on points $x$ of Minkowski space-time situated on $\Sigma$, and on parameter $\tau$. The "Lagrangian" is given by 
 \begin{align}
 	{L}(\overline{\phi},\phi )&=-\overline{\phi}\frac{1}{2}\gamma^0\overset{\leftrightarrow}{\partial}_\tau\phi+ \beta(0,\vec{x})\Bigl(\sum_i\mu_ij^{0}_i \nonumber\\&  
 	- { u}_0(0,\vec{x}) m\overline{\phi}\phi -{ u}^k(0,\vec{x})\overline{\phi}i\gamma^j\overset{\leftarrow}{\partial}_j\gamma^0\frac{1}{4}\gamma_{k}\phi \nonumber\\&
 	+{ u}^k(0,\vec{x})\overline{\phi}\frac{1}{4}\gamma_{k}\gamma^0 i\gamma^j\overset{\rightarrow}{\partial}_j\phi -\overline{\phi}\frac{i}{4}\gamma_{0}\overset{\leftrightarrow}{\partial}_{j}\phi { u}^{j}(0,\vec{x}) \nonumber\\&-{ u}^0(0,\vec{x})\overline{\phi}i\gamma^j\overset{\leftarrow}{\partial}_j\frac{1}{2}\phi 
 	+{ u}^0(0,\vec{x})\overline{\phi}\frac{1}{2} i\gamma^j\overset{\rightarrow}{\partial}_j\phi \Bigr)\label{Z1u} 
 \end{align}
It is assumed here that the operator of $i$ - th conserved charge has the form 
$$
\hat j^\mu_i = \hat{\bar{\Psi}} {\bf t}^\mu_i \hat{\Psi} 
$$
where ${\bf t}^\mu_i$ is an operator acting, in particular, in internal space, as well as acting on spinor index  (in the simple particular case of electric current it is equal to $\gamma^\mu$). Then the corresponding current in the above lagrangian is
$$
j^\mu_i = \bar{\phi} {\bf t}^\mu_i \phi 
$$
The four - velocity entering the above expression ${u}(0,\vec{x})$ coincides with $u(x)$ at $x \in \Sigma$. Here the initial moment in time is set to $0$: surface $\Sigma$ initially was taken as the hyperplane $t = 0$. The same refers to ${\beta}(0,\vec{x})$ - it coincides with $\beta(x)$ at $x \in \Sigma$. Both these functions do not depend on $\tau$.

One can represent $Z[n(x),u(x),\beta(x),\mu_i(x)] = {\cal Z}[n(x),u(x),\beta(x),\mu_i(x), - i]$, where
\begin{eqnarray}
&& {\cal 	Z}[n,u,\beta(x),\mu_i(x), h]\nonumber\\ && ={\rm Tr}\,{\rm exp}\,(-i h {\cal R}[\beta(x),\mu_i(x),u^\mu(x),{\hat{\Psi}}(x),\bar{{\hat{\Psi}}}(x)])\nonumber\\& &=  \int D\bar{\phi} D\phi \, e^{i\int d \Sigma \int_0^h d w \, \mathcal{L}(\overline{\phi},\phi )}
\end{eqnarray}
Integration in the exponent of the above expression is over the piece of $\Sigma \otimes R$ that consists of points $(x,w)$ with $w \in (0,h)$. The fields $\phi(x,w)$ and  $\bar{\phi}(x,w)$ are now the functions of $x\in \Sigma$ and $w \in R$.  The new lagrangian is given by
 \begin{align}
	\mathcal{L}(\overline{\phi},\phi )&=\overline{\phi}\frac{i}{2}\gamma^0\overset{\leftrightarrow}{\partial}_w\phi+ {\beta}(0,\vec{x})\Bigl(\sum_i\mu_ij^{0}_i   
	\nonumber\\&- u_0(0,\vec{x}) m\overline{\phi}\phi -u^k(0,\vec{x})\overline{\phi}i\gamma^j\overset{\leftarrow}{\partial}_j\gamma^0\frac{1}{4}\gamma_{k}\phi \nonumber\\&
	+u^k(0,\vec{x})\overline{\phi}\frac{1}{4}\gamma_{k}\gamma^0 i\gamma^j\overset{\rightarrow}{\partial}_j\phi -\overline{\phi}\frac{i}{4}\gamma_{0}\overset{\leftrightarrow}{\partial}_{j}\phi u^{j}(0,\vec{x}) \nonumber\\&-u^0(0,\vec{x})\overline{\phi}i\gamma^j\overset{\leftarrow}{\partial}_j\frac{1}{2}\phi 
	+u^0(0,\vec{x})\overline{\phi}\frac{1}{2} i\gamma^j\overset{\rightarrow}{\partial}_j\phi \Bigr)\label{Z2u} 
\end{align}
Now instead of $\Sigma \otimes R$ we restore Minkowski space with the time variable related to $w$ via rescaling
\begin{equation}
	t = w\,{\mathfrak B}(\vec{x}) \label{rescaling}
\end{equation}
 with certain scaling function $\mathfrak B$ of spatial coordinates to be specified below. The new fields  $\psi(t,\vec{x})$ and  $\bar{\psi}(t,\vec{x})$ are defined as 
$$
\psi(t,\vec{x}) = \phi(t/{\mathfrak B}(\vec{x}),\vec{x}), \quad \bar{\psi}(t,\vec{x}) =  \bar{\phi}(t/{\mathfrak B}(\vec{x}),\vec{x}),
$$
 where $\vec{x}\in \Sigma$. In terms of these fields we have 
\begin{eqnarray}
	{\cal 	Z}[n,u,\beta(x),h]&=&\int D\bar{\psi} D\psi \, e^{i\int d^4 x   \,\mathcal{L}(\overline{\psi},\psi )}\label{calZ}
\end{eqnarray}
The integration in the exponent here is along 
$4D$ shell having as one of its boundaries hyperplane $\Sigma$, its second boundary is (in general, curved) hypersurface depending on function $\mathfrak B$ introduced above: $\Sigma_h = \{(h {\mathfrak B}(\vec{x}),\vec{x})|\vec{x} \in \Sigma\}$. At the same time the lagrangian is given by:
\begin{align}
	\mathcal{L}(\overline{\psi},\psi )&=\overline{\psi}\frac{i}{2}\gamma^0\overset{\leftrightarrow}{\partial}_0\psi+ \sum_i\mu_ij^{0}_i   
	- {\mathfrak U}_0 m\overline{\psi}\psi \nonumber\\&-{\mathfrak U}^k\overline{\psi}i\gamma^j\overset{\leftarrow}{\partial}_j\gamma^0\frac{1}{4}\gamma_{k}\psi 
	+{\mathfrak U}^k\overline{\psi}\frac{1}{4}\gamma_{k}\gamma^0 i\gamma^j\overset{\rightarrow}{\partial}_j\psi \nonumber\\&-\overline{\psi}\frac{i}{4}\gamma_{0}\overset{\leftrightarrow}{\partial}_{j}\psi {\mathfrak U}^{j} -{\mathfrak U}^0\overline{\psi}i\gamma^j\overset{\leftarrow}{\partial}_j\frac{1}{2}\psi \nonumber\\&
	+{\mathfrak U}^0\overline{\psi}\frac{1}{2} i\gamma^j\overset{\rightarrow}{\partial}_j\psi \label{Z1u} 
\end{align}
In this expression we introduce four - vector
\begin{equation}
	{\mathfrak U}(\vec{x}) = \frac{\beta(0,\vec{x})}{{\mathfrak B}(\vec{x})} u(0, \vec{x})\label{mU}
\end{equation}

The anti - periodic boundary conditions are implied: 
$$
\psi(  {\mathfrak B}(\vec{x}) h, \vec{x}) = - \psi(0,\vec{x}), \quad \bar{\psi}(  {\mathfrak B}(\vec{x}) h, \vec{x}) = - \bar{\psi}(0,\vec{x})
$$
Eq. (\ref{Z1u}) is the effective lagrangian of the system that remains in global equilibrium for motion with 4 - velocity $u$. {\it Notice that although we define here the partition function in Minkowski space - time, function ${\mathfrak U}$ entering the effective lagrangian remains the function of the spatial components of $x$ only. } The types of fields ${u}(t,x)$ that are allowed for the equilibrium include motion with constant velocity (which is reduced to the system at rest in the corresponding reference frame), the rigid rotation (see Appendix \ref{rigid}) and "accelerated" motion (see Appendix \ref{accelerate}). Notice that the latter case does not refer to the true accelerated system, but rather to the system that is accelerated at the initial moment.

The expression for the effective lagrangian may be rewritten as
\begin{align}
	\mathcal{L}(\overline{\psi},\psi )  =&
	{\mathfrak U}^0\overline{\psi}(\frac{1}{2} i\gamma^\mu\overset{\leftrightarrow}{\partial}_\mu-m)\psi  +(1-{\mathfrak U}_0)\overline{\psi}\frac{i}{2}\gamma^0\overset{\leftrightarrow}{\partial}_0\psi\nonumber\\&+ \sum_i\mu_ij^{0}_i   
	-{\mathfrak U}^k\overline{\psi}i\gamma^j\overset{\leftarrow}{\partial}_j\gamma^0\frac{1}{4}\gamma_{k}\psi \nonumber\\
	&+{\mathfrak U}^k\overline{\psi}\frac{1}{4}\gamma_{k}\gamma^0 i\gamma^j\overset{\rightarrow}{\partial}_j\psi -\overline{\psi}\frac{i}{4}\gamma_{0}\overset{\leftrightarrow}{\partial}_{j}\psi {\mathfrak U}^{j} \nonumber\\
	&={\mathfrak U}^0\overline{\psi}(\frac{1}{2} i\gamma^\mu\overset{\leftrightarrow}{\partial}_\mu-m)\psi  +(1-{\mathfrak U}_0)\overline{\psi}\frac{i}{2}\gamma^0\overset{\leftrightarrow}{\partial}_0\psi\nonumber\\&+ \sum_i\mu_ij^{0}_i   
	\nonumber\\
	&~~{+  {\mathfrak U}_k(\overline{\psi}\gamma^0(\frac{i}{8}[\gamma^{j},\gamma^k](\overset{\leftarrow}{\partial}_j+\partial_j) 
		- \frac{i}{2}\overset{\leftrightarrow}{\partial^k})\psi)}
\end{align}

The choice of function ${\mathfrak B}(\vec{x})$ is free. We mention here the two particular cases, which lead to the two different interpretations of the meaning of the four - vector $\mathfrak U$:

\begin{enumerate}
	\item ${\mathfrak B}(\vec{x}) = \beta(0,\vec{x})$. In this case 
	\begin{equation}
		{\mathfrak U}^\mu(\vec{x}) =  u^\mu(0, \vec{x})\label{mU1}
	\end{equation}
Then vector field $\mathfrak U$ may be interpreted as the four - velocity distribution at the initial moment. 

\item ${\mathfrak B}(\vec{x}) = \beta(0,\vec{x}) u^0(0,\vec{x})$. In this case 
\begin{equation}
	{\mathfrak U}^\mu(\vec{x}) =  \frac{u^\mu(0, \vec{x})}{u^0(0, \vec{x})}\label{mU2}
\end{equation}
Then vector field $\mathfrak U$ cannot be interpreted as four - velocity of macroscopic motion. However, this case results in relatively simple effective expression for effective lagrangian for the important particular case of rigid rotation. 

With this choice we obtain
\begin{align}
	\mathcal{L}(\overline{\psi},\psi )  
	&=\overline{\psi}(\frac{1}{2} i\gamma^\mu\overset{\leftrightarrow}{\partial}_\mu-m)\psi  + \sum_i\mu_ij^{0}_i   
	\nonumber\\
	&~~{+  {\mathfrak U}_k(\overline{\psi}\gamma^0(\frac{i}{8}[\gamma^{j},\gamma^k](\overset{\leftarrow}{\partial}_j+\partial_j) 
		- \frac{i}{2}\overset{\leftrightarrow}{\partial^k})\psi)}
\end{align}
	
\end{enumerate}

\section{Dirac fermions with non-Abelian gauge interactions}

\subsection{Pure gauge field action}

The treatment of the Zubarev statistical operator in the previous section was restricted to the non-interacting case. We now couple the fermions to non-Abelian gauge bosons. This amounts to the replacement of the ordinary derivative by a covariant one $\partial_{\mu}\to D_{\mu}=\partial_{\mu}-igA_{\mu}$ (in fundamental representation). Doing so for the initial Lagrangian, the time derivative in the definition of the canonically conjugate momentum and the time derivative in the Heisenberg equation of motion implies the validity of the replacement $\partial_{\mu}\to D_{\mu}$ throughout the rest of the previous section.\par
In addition we have to account for the energy momentum tensor of the non-Abelian gauge fields themselves. To this end we start out from the Yang-Mills action. We employ the temporal gauge, which allows for a compact treatment of the canonical quantization of the theory. 

The pure Yang-Mills action  (in the presence of gravitational field) is given by

\begin{align}
	\nonumber S_B&=\int d^4x \sqrt{-g}(-\frac{1}{2}\tr(F_{\mu\nu}F^{\mu\nu}))\nonumber\\&=\int d^4x \sqrt{-g}(-\frac{1}{4}F^a_{\mu\nu}F^{a\mu\nu}).
\end{align}

Index $a$ represents the sum over Lie algebra generators $T^a$ of the gauge group, which satisfy relations

\begin{align}
	\tr(T^aT^b)=\frac{1}{2}\delta^{ab},\,\,\,\, {[T^a,T^b]=if^{cab}T^c}
\end{align}
with the structure constants $f^{abc}$.  The gauge field $A_{\mu}$ and the field strength $F_{\mu\nu}$ have the expansions
\begin{align}
&	A_{\mu}=A^a_{\mu}T^a,\,\,\,\, F_{\mu\nu}=F^a_{\mu\nu}T^a,\,\,\,\, \nonumber\\&{F_{\mu\nu}=\partial_{\mu}A_{\nu}-\partial_{\nu}A_{\mu}-i g [A_{\mu},A_{\nu}]}
\nonumber\\&F^a_{\mu\nu}=\partial_{\mu}A^a_{\nu}-\partial_{\nu}A^a_{\mu}+gf^{abc}A^b_{\mu}A^c_{\nu}.
\end{align}

The metric determinant can be represented in the form

\begin{align}
	g=\frac{1}{4!}\epsilon^{\alpha\beta\gamma\delta}\epsilon^{\mu\nu\rho\sigma}g_{\alpha\mu}g_{\beta\nu}g_{\gamma\rho}g_{\delta\sigma}
\end{align}

with the Levi-Civita tensor density $\epsilon^{\alpha\beta\gamma\delta}$ defined by the convention $\epsilon^{0123}=1$. In the following we will need expression for the action in inertial reference frame with $g = 1$ except for the case, when the derivative with respect to $g$ is needed in order to calculate stress - energy tensor.

\subsection{The theory in temporal gauge}
\label{temporal}

We introduce quantized nonabelian gauge field $\hat{\cal A}^{a}_\mu$. In temporal gauge $\hat{\cal A}^a_0 = 0$ the gauge is not fixed completely (the gauge freedom remains related to the gauge transformations that depend on spatial coordinates only). Nevertheless, the canonical quantization may be performed easily resulting in relatively simple expression for the Hamiltonian, and the canonical commutation relations. For the details see \cite{ROSSI1980109} and Appendix \ref{path}. The Hamiltonian and canonical commutation relations are to be supplemented by the Gauss constraint:
\begin{equation}
\hat{G}^a = 	D_\mu \hat{\cal F}^{\mu 0 } +\hat J^{0} = 0,\label{constr2}
\end{equation}
where $\hat{\cal F}^a_{\mu\nu}=\partial_{\mu}\hat{\cal A}^a_{\nu}-\partial_{\nu}\hat{\cal A}^a_{\mu}+gf^{abc}\hat{\cal A}^b_{\mu}\hat{\cal A}^c_{\nu}$ is non - Abelian field strength,  $\hat J^{a,\nu}= g\bar{\Psi} \gamma^\nu T^a \Psi$ is the color current (here $T$ is generator of the gauge group). {The Gauss constraint should be understood in the weak sense, i.e.
$$
\hat G^a \ket{\Phi}=0
$$
for any physical vector of Hilbert space. In turn, this condition is nothing but the requirement that the wave functionals are invariant under the residual gauge transformations (depending on spatial coordinates) - see Appendix \ref{path}.}

The quantum average of any operator $\cal O$ can be written as 
\begin{eqnarray}
	\langle {\cal O} \rangle & = &  {\rm Tr}\, P\, e^{i \int d^4 x \,\sqrt{-g}\, \hat{\cal H} }{\cal O} e^{-i \int d^4 x \,\sqrt{-g}\, \hat{\cal H}} \,\hat{\rho}
\end{eqnarray}
Here $\hat \rho$ is initial density matrix, we denote by $P$ ordering of the operators standing inside the trace along Keldysh contour: it starts from $t = 0$, goes to plus infinity, and returns back. More precisely, this ordered product looks as  
$$
P\, e^{i \int d^4 x \,\sqrt{-g}\, \hat{\cal H}} \equiv e^{i \int_{\Sigma_0} d\Sigma_\mu (d x^\mu/d\sigma) \delta \sigma \hat{\cal H}} \ldots $$ $$ \ldots e^{i \int_{\Sigma_\sigma} d\Sigma_\mu (d x^\mu/d\sigma) \delta \sigma \hat{\cal H}}\ldots
$$
Here we again mention the foliation of space - time into the slices $\Sigma_\sigma$. The Hamiltonian density (in temporal gauge) for the theory that contains both fermions and gauge fields is given by  
$$
\hat{\cal H} = {\cal H}_\psi[\hat{\cal A}, \Psi] +\frac{1}{2} \, \Big(\hat{\Pi}_{i} \hat{\Pi}_{i} +   \hat{\cal B}_{i} \hat{\cal B}_{i}\Big)
$$
(sum is over $i = 1,2,3$ is assumed).
Here ${\cal H}_\psi[\hat{\cal A}, \Psi]$ is the Hamiltonian density for the fermion operator field $\Psi$ interacting with gauge field $\hat{\cal A}$. (Nonabelian) magnetic field strength is 
$\hat{\cal B}^a_i=-\frac{1}{2}\epsilon_{ijk}\hat{\cal F}^a_{jk}$. Operator $\hat{\Pi}_{i}$ is momentum, canonically conjugated to the field $\hat{\cal A}$. These fields obey the commutation relations
{$$
[\hat{\Pi}^a_{i}(x),\hat{\cal A}^{jb}(y)] =  -i \delta^{ab}\delta^j_i \delta^{(3)}_{\Sigma}(x-y)\Big|_{x, y \in \Sigma_\sigma}
$$}
Notice that on the level of classical theory the canonically conjugate momentum is equal to electric field strength $\Pi_{i} = E_{i}=F_{i0}$, which means that in quantum theory this equality remains being placed inside the functional integral. In the other words, we have operator relation
\begin{equation}
	\hat\Pi_{i} = \hat {\cal E}_{i}\label{constr1}
\end{equation}
where $\hat{\cal E}_i = \hat{\cal F}_{i0}$.
Besides, the {Gauss constraint 
$$
\hat G^a\ket{\Phi} = (D_\mu \hat{\cal F}^{a\mu 0 } + \hat{J}^{a0})\ket{\Phi} = 0
$$
 is to be imposed for any physical state $\ket{\Phi}$. Being imposed at the initial moment of evolution in time this condition is not changed. }

\subsection{Stress - energy tensor}

In classical theory the Lagrangian may be written as
\begin{align}
	\mathcal{L}&=-\frac{1}{4}F^a_{\mu\nu}F^{a\,\mu\nu}\rvert_{A^a_0=0}=-\frac{1}{4}F^a_{ij}F^{aij}-\frac{1}{2}\partial_0A^a_i\partial_0A^{ai}\nonumber\\&=-\frac{1}{2}B^a_iB^a_i+\frac{1}{2}\partial_0A^a_i\partial_0A^{a}_i
\end{align}
with $B^a_i=-\frac{1}{2}\epsilon_{ijk}F^a_{jk}$ together with the Gauss law constraint. This ansatz leaves the residual gauge freedom, which may be fixed completely by requiring $\partial^iA^a_i=0$ on a spacelike hypersurface $\Sigma$. (In quantum theory then the ghosts will appear.) 

The canonical energy momentum tensor of Yang-Mills theory in covariant notation has the form
\begin{align}
	T^{can}_{\mu\nu}=\frac{\partial\mathcal{L}}{\partial (D^{ac\mu}A^c_{\rho})}D^{ab}_{\nu}A^b_{\rho}-g_{\mu\nu}\mathcal{L}=F^{a\rho}_{\mu}D^{ac}_{\nu}A^c_{\rho}-g_{\mu\nu}\mathcal{L},
\end{align}
which is different from the Belinfante-Rosenfeld energy momentum tensor due to the appearance of the spin current
\begin{align}
	S^{ac}_{\mu\nu\rho}=-F^a_{\lambda\mu}A^c_{\nu}+F^a_{\nu\mu}A^c_{\lambda}.
\end{align}

The Belinfante-Rosenfeld energy momentum tensor then coincides modulo equations of motion with the gravitational energy momentum tensor

\begin{align}
	T_{\mu\nu}(x)&=\frac{2}{\sqrt{-g}}\frac{\delta S_B}{\delta g^{\mu\nu}(x)}\nonumber\\&=F^{a\rho}_{\,\,\,\,\, \mu}(x)F^a_{\nu\rho}(x)+\frac{1}{4}g_{\mu\nu}(x)F^a_{\rho\sigma}(x)F^{a\rho\sigma}(x).\nonumber
\end{align}

\subsection{Zubarev statistical operator}

The full Zubarev statistical operator including interactions reads
\begin{align}
	\log\hat \rho =-\alpha -\int d\Sigma\beta n^{\mu}g_{\mu\nu}((\hat T^{}_F)^{\nu\rho}+(\hat T^{}_B)^{\nu\rho})u_{\rho}-\sum_i\hat j^{\nu}_i)
\end{align}
where the  energy momentum tensor $\hat{T}^{}_F$ of fermion fields contains gauge covariant derivatives and $\hat T^{}_B$ is the energy momentum tensor of the gauge fields. We will calculate it in the following after canonically quantizing the temporal gauge simplified Lagrangian and subsequently the Gauss constraint.\newline

As above we assume that the foliation of space - time is given by hyperplanes $t = const$, and, correspondingly,  $n^{\mu}=(1,0,0,0)$. The canonically conjugate momentum fields in temporal gauge of classical theory are

\begin{align}
	\Pi^{ai}=\frac{\partial\mathcal{L}}{\partial (\partial_0A^a_i)}=\partial_0A^a_i
\end{align}

and imply the canonical commutation relations for the corresponding operators

\begin{align}
&	[\hat{\cal A}_i^a(\vec{x}),\hat{\cal A}_j^b(\vec{y})]=[\hat{\Pi}^a_i(\vec{x}),{\hat\Pi}_j^b(\vec{y})]=0,\nonumber\\&\,\,\,\, [\hat{\cal A}_i^a(\vec{x}),{\hat\Pi}_j^b(\vec{y})]={-}i\delta_{ij}\delta^{ab}\delta^3(\vec{x}-\vec{y}).
\end{align}

The classical Hamiltonian is given by

\begin{align}
&	\mathcal{H}=\Pi^{ai}\partial_0A^a_i-\mathcal{L}=\frac{1}{4}F^a_{ij}F^{aij}+\frac{1}{2}\Pi^{ai}\Pi^{ai}\nonumber\\&=\frac{1}{2}B^a_iB^a_i+\frac{1}{2} \Pi^{ai} \Pi^{ai}.
\end{align}

{At the quantum level we have $G^a\rvert \Phi \rangle =0$ for any physical state $\rvert \Phi \rangle$ of the Hilbert space. }

In quantum theory we substitute in the above expression for the Hamiltonian the operator fields $\hat{\Pi}$ and $\hat{\cal A}$ instead of $\Pi$ and $A$. 

The classical energy momentum tensor in temporal gauge may be represented by
\begin{align}
	\nonumber T_{\mu\nu}=&\delta_{\mu}^0\delta_{\nu}^0(\frac{1}{2} \Pi^{ai} \Pi^{ai}+\frac{1}{4}F^a_{ij}F^{aij})\nonumber\\&+(\delta_{\mu}^j\delta_{\nu}^0+\delta_{\mu}^0\delta_{\nu}^j) \Pi^{ai}F^a_{ji}\nonumber\\
	\nonumber &+\frac{1}{2}(\delta_{\mu}^i\delta_{\nu}^j+\delta_{\mu}^j\delta_{\nu}^i)(F^a_{ik}F^a_{jk}-\frac{1}{4}\delta_{ij}F^a_{kl}F^{akl}\nonumber\\&+\frac{1}{2}\delta_{ij}\Pi^{ak}\Pi^{ak}-\Pi^{ai}\Pi^{aj})\nonumber\\
	\nonumber =&\delta_{\mu}^0\delta_{\nu}^0(\frac{1}{2}\Pi^{ai} \Pi^{ai}+\frac{1}{2}B^a_iB^a_i)\nonumber\\&+(\delta_{\mu}^j\delta_{\nu}^0+\delta_{\mu}^0\delta_{\nu}^j)\epsilon_{ijk}\Pi^{ai}B^a_k\nonumber\\
	&+{\frac{1}{2}}(\delta_{\mu}^i\delta_{\nu}^j+\delta_{\mu}^j\delta_{\nu}^i)(\frac{1}{2}\delta_{ij}B^a_kB^a_k-B^a_iB^a_j\nonumber\\&+\frac{1}{2}\delta_{ij}\Pi^{ak}\Pi^{ak}-\Pi^{ai}\Pi^{aj}).
\end{align}

Notice that we did not fix the gauge completely, the gauge freedom remains on each $\Sigma$.

\subsection{Effective functional integral representation of the theory  }

First of all, we represent statistical operator as 
$$
\rho = e^{-\alpha} \lim_{N\to \infty}\Pi_{s}e^{-{\cal R}[\beta(x),u^\mu(x),\hat\Psi(x),\bar{\hat\Psi}(x),\hat{\cal A}(x),\hat\Pi(x)]\delta(N)},$$$$ \, \delta(N)=1/N, \,\, s=0,1,\ldots,N-1
$$
Next, we use relations given in Appendix \ref{path}, and insert unities from Eqs. (\ref{AP}) and (\ref{PP}) between each two multipliers in this product, and arrive at the expression for  the partition function 
\begin{eqnarray}
	&&Z[n(x),u(x),\beta(x),\mu_i(x)]= e^{\alpha}\nonumber\\&&
	= \int D\bar{\phi} D\phi DA_i D\Pi\, e^{\int_0^1 d \tau \int d\Sigma {L}(\overline{\phi},\phi,A,\Pi )}
\end{eqnarray}
Integration in this expression is over the spatial components $A_i$ of gauge field.
Here we fix one particular hyperplane  $\Sigma$ in Minkowski space (corresponding to $n=(1,0,0,0)$). Now $A(x,\tau)$, $\phi(x,\tau)$ and $\tilde{\phi}(x,\tau)$ are independent fields depending on points $x$ of Minkowski space-time situated on $\Sigma$, and on parameter $\tau$, while $\bar{\phi}=\tilde{\phi}\gamma^0$; "Lagrangian" is given by 
\begin{align}
	{L}(\overline{\phi},\phi,A,\Pi )&=-\overline{\phi}\frac{1}{2}\gamma^0\overset{\leftrightarrow}{\partial}_\tau\phi+ {\beta}(0,\vec{x})\Bigl(\sum_i\mu_ij^{0}_i   
	\nonumber\\&- {u}_0(0,\vec{x}) m\overline{\phi}\phi -{u}^k(0,\vec{x})\overline{\phi}i\gamma^j\overset{\leftarrow}{D}_j\gamma^0\frac{1}{4}\gamma_{k}\phi \nonumber\\&
	+{u}^k(0,\vec{x})\overline{\phi}\frac{1}{4}\gamma_{k}\gamma^0 i\gamma^j\overset{\rightarrow}{D}_j\phi -\overline{\phi}\frac{i}{4}\gamma_{0}\overset{\leftrightarrow}{D}_{j}\phi {u}^{j}(0,\vec{x}) \nonumber\\&-{u}^0(0,\vec{x})\overline{\phi}i\gamma^j\overset{\leftarrow}{D}_j\frac{1}{2}\phi 	+{u}^0(0,\vec{x})\overline{\phi}\frac{1}{2} i\gamma^j\overset{\rightarrow}{D}_j\phi \Bigr)\nonumber\\& 
	+\Pi^{ak}i\frac{{\partial}}{\partial \tau} A^a_{k} -{\beta}(0,\vec{x})  T_{0\mu}[A,\Pi] {u}^\mu(0,\vec{x}) 
\end{align}
The covariant derivatives here act on $\phi$ and $\bar{\phi}$ only. $T_{cd}[A,\Pi]$ is stress - energy tensor of the gauge field. 

Next, following analogy with the above considered case of free fermions, we can also represent $Z[n(x),u(x),\beta(x),\mu_i(x)] = {\cal Z}[n(x),u(x),\beta(x),\mu_i(x) -i]$, which brings us from Euclidean space - time of statistical field theory to Minkowski space - time of effective quantum field theory of substance in the state of macroscopic motion 
\begin{eqnarray}
	{\cal 	Z}[n,u,\beta(x),\mu_i(x),h]&=&\int D\bar{\psi} D\psi DA_i D\Pi  \, e^{i\int  d^4 x \, \mathcal{L}(\overline{\psi},\psi ,A, \Pi)}\nonumber
\end{eqnarray}
Here field $\psi$ is defined in Minkowski space - time. Integration is over $4D$ shell having as one of its boundaries surface $\Sigma$, its second boundary is the geometric place of points $\{ x^\mu + n^\mu {\mathfrak B}(\vec{x}) h| x\in \Sigma \}$. Rescaling function ${\mathfrak B}(\vec{x})$ may be chosen arbitrary. The resulting $3+1$ D lagrangian is  
\begin{align}
	{\mathcal L}(\overline{\psi},\psi,A,\Pi )&={\mathfrak U}^0\overline{\psi}(\frac{1}{2} i\gamma^\mu\overset{\leftrightarrow}{D}_\mu-m)\psi \nonumber\\& +(1-{\mathfrak U}_0)\overline{\psi}\frac{i}{2}\gamma^0\overset{\leftrightarrow}{D}_0\psi+ \sum_i\mu_ij^{0}_i   
	\nonumber\\
	&~~{+  {\mathfrak U}_k(\overline{\psi}\gamma^0(\frac{i}{8}[\gamma^{j},\gamma^k](\overset{\leftarrow}{D}_j+D_j) 
		-\frac{i}{2} \overset{\leftrightarrow}{D^k})\psi)}\nonumber\\& 
	+\Pi^{ak}\frac{{\partial}}{\partial t} A^a_{k} -  T_{0\mu}[A,\Pi] {\mathfrak U}^\mu 
\end{align}
Here vector field $\mathfrak U$ is given by Eq. \ref{mU}. The two specially important particular cases for the choice of function ${\mathfrak B}(\vec{x})$ are given by Eqs. (\ref{mU1}) and (\ref{mU2}).

We can rewrite the last expression as
{(note that the derivatives act on \(\overline{\psi},\psi\), but not on \({\mathfrak U}_\mu\))}
\begin{align}
	\mathcal{L}(\overline{\psi},\psi,A,\Pi)
	=& {\mathfrak U}^0(\overline{\psi}(\gamma^\mu\frac i2\overset{\leftrightarrow}{D}_\mu-m)\psi) \nonumber\\& 
		+(1-{\mathfrak U}_0)(\overline{\psi}\gamma^0\frac{i}{2}\overset{\leftrightarrow}{D}_0\psi) + \sum_i\mu_ij^{0}_i + \nonumber\\
	&+ { {\mathfrak U}_k(\overline{\psi}\gamma^0(\frac{i}{8}[\gamma^{j},\gamma^k](\overset{\leftarrow}{D}_j+D_j) 
		-\frac{i}{2} \overset{\leftrightarrow}{D^k})\psi)} + \nonumber\\
	&+\Pi^{ak} { E}^a_{k} -  {\mathfrak U}_0 (\frac{1}{2}\Pi^{ai}\Pi^{ai}+\frac{1}{2}{ B}^a_i{ B}^a_i)\nonumber\\& -  \epsilon_{ijk}\Pi^{ai}{ B}^a_k {\mathfrak U}^j  
\end{align}
with ${ E}^a_{k} = \partial_t A^a_k$ and $B^a_k = -\frac{1}{2}\epsilon_{kij}F^{aij}$. Integration over $\Pi$ may be performed, and we arrive at 
\begin{equation}
	{\cal 	Z}[n(x),u(x),\beta(x),\mu_i(x),h]=\int D\bar{\psi} D\psi DA_i  \, e^{i\int  d^4 x \, \mathcal{L}(\overline{\psi},\psi,A)}\label{Z4}
\end{equation} 
with
\begin{align}
	\mathcal{L}(\overline{\psi},\psi,A)
	=& {\mathfrak U}^0(\overline{\psi}(\gamma^\mu\frac i2\overset{\leftrightarrow}{D}_\mu-m)\psi)  
	\nonumber\\&	+(1-{\mathfrak U}_0)(\overline{\psi}\gamma^0\frac{i}{2}\overset{\leftrightarrow}{D}_0\psi) + \sum_i\mu_ij^{0}_i + \nonumber\\
	&+ { {\mathfrak U}_k(\overline{\psi}\gamma^0(\frac{i}{8}[\gamma^{j},\gamma^k](\overset{\leftarrow}{D}_j+D_j) 		-\frac{i}{2} \overset{\leftrightarrow}{D^k})\psi)} + \nonumber\\
	&+ \frac{1}{2 {\mathfrak U}_0}\tilde{E}^{ai}\tilde{E}^{ai}-\frac{{\mathfrak U}_0}{2}B^a_iB^a_i,\label{rez1}
\end{align}
where
$$
\tilde{E}^{a}_{i} = E^{a}_i - \epsilon_{ijk} B^a_k {\mathfrak U}^j.
$$
	The given above lagrangian simplifies for the choice of function $\mathfrak B$ given by Eq. (\ref{mU2}). Then the first line is given by usual Lagrangian for the Dirac field;
	the second line contains the term with chemical potentials multiplied by the corresponding charge densities, while the term proportional to $1-{\mathfrak U}^0$ disappears completely; the third line (after integration by parts) may be reduced to vorticity multiplied by spin of the Dirac field; the last line contains the modified lagrangian of gauge field.

{\it Recall that the theory given by Eqs. (\ref{Z4}) and (\ref{rez1}) is to be supplemented by Gauss constraint, i.e. the requirement that the Wave functionals of gauge field $\Phi[A]$ are invariant under spatial dependent gauge transformations (the gauge freedom that remains after the fixing of temporal gauge).  }  

\subsection{Gauge invariance restoration}

In order to avoid dealing with Gauss constraint one can bring the effective theory to the gauge invariant form. For this purpose we use representation
\begin{equation}
	E^a_i = F^a_{\mu i}n^\mu, \quad B^a_i = - \frac{1}{2}\epsilon_{\mu ijk}F^{ajk}n^\mu
\end{equation}
This allows to represent the above expression for effective lagrangian as
\begin{align}
	\mathcal{L}(\overline{\psi},\psi,A)
	=& {\mathfrak U}_0(\overline{\psi}(\gamma^\mu\frac i2\overset{\leftrightarrow}{D}_\mu-m)\psi)  
	\nonumber\\&	+(1-{\mathfrak U}_0)(\overline{\psi}\gamma^0\frac{i}{2}\overset{\leftrightarrow}{D}_0\psi) + \sum_i\mu_ij^{0}_i + \nonumber\\
	&{+ {\mathfrak U}_k(\overline{\psi}\gamma^0(\frac{i}{8}[\gamma^{j},\gamma^k](\overset{\leftarrow}{D}_j+D_j) 
		-\frac{i}{2} \overset{\leftrightarrow}{D^k})\psi)} + \nonumber\\
	&- {\frac{1}{4 {\mathfrak U}_0}\Bigl({F}^{a\mu\nu}{F}^{a}_{\mu\nu}+({\mathfrak U}_\mu {\mathfrak U}^\mu-1){F}^{aij}{F}^{a}_{ij}\Bigr)}\nonumber\\& - \frac{1}{8 {\mathfrak U}_0}(\epsilon_{lmn}{F}^{amn}{\mathfrak U}^l)(\epsilon_{ijk}{F}^{ajk}{\mathfrak U}^i)\nonumber\\
	&-\frac{1}{ {\mathfrak U}_0}{F}^{a}_{i0}{F}^{aij}{\mathfrak U}_j.\label{finL4}
\end{align} 
The partition function of effective theory is given by Eq. (\ref{Z4}), in which integration over spatial components $A_i$ of gauge field is substituted by integration over all space - time components $A_\mu$. From now on we can forget about the derivation performed in temporal gauge, and work directly with effective gauge invariant theory. Any gauge may be fixed in this theory using the Faddeev - Popov procedure, or the lattice simulations may be performed in the manifestly gauge invariant model after Wick rotation and lattice discretization.

\section{Discussion}

\subsection{Types of macroscopic motion admitted for thermal equilibrium}

We started from the conventional Zubarev statistical operator defined for the arbitrary foliation of space - time to space - like surfaces $\Sigma_\sigma$. Macroscopic velocity $u(x)$ appears in this approach naturally, and it appears that the following types of macroscopic motion are possible in thermodynamic equilibrium \cite{Zubarev1979}:

\begin{enumerate}
	\item 
	
	Motion with constant four - velocity $u(x) = {\rm const}$. Correspondingly, in this case temperature, and chemical potential are constant as functions of time.
	
	\item
	
	Rigid rotation with constant angular velocity $\omega$ (see Appendix \ref{rigid}). In this case temperature becomes function of space point. The whole theory becomes ill - defined at the distances from the rotation axis larger than $1/\omega$. This means that we can use the Zubarev statistical operator for the case, when $\xi \omega < 1$, where $\xi$ is the size of the considered system. This admits, in particular, the possibility of rotation with relativistic velocities. 
	
	If rotation is along the $z$ - axis, then we have 
	$$
	u(x) = \frac{1}{\sqrt{1-\omega^2(x^2 + y^2)}}(1,-y\omega,x\omega,0)
	$$
	and
	$$
	\beta(x) = \beta_0 \sqrt{1-\omega^2(x^2+y^2)}, \quad b = (\beta_0,0,0,0)
	$$ 
	with constant $\beta_0$ of dimension of inverse temperature.
	
	\item 
	
	The third type of motion admitted for the thermodynamic equilibrium is typically referred to as the accelerated motion (for the details see Appendix \ref{accelerate}). However, this identification is not exact. Actually, the corresponding "acceleration" $a$ appears as the thermodynamically conjugated quantity to the boost operator. The accelerated motion itself appears only at the initial time of motion. At the later times it is reduced to the other type of motion, moreover, the interpretation of the theory in terms of the four - velocity $u(x)$ becomes ill - defined at time $t>1/a$.  However, the spatial size of the corresponding system is not limited. 
	
	In the case, when acceleration is along axis $x$, we have: 
	$$
	u(x) = \frac{1}{\sqrt{(1+ax)^2 - a^2 t^2}}(1+ax,at,0,0)
	$$
	and
	$$
	\beta(x) = \beta_0 \sqrt{(1+ax)^2 -a^2 t^2}
	$$

	\item 
	
	The combination of the three above types of motion is also admitted for thermal equilibrium. 
	
\end{enumerate}


\subsection{Partition function for Zubarev statistical operator in terms of functional integral}

Statistical partition function of equilibrium system of fermions interacting with the non - Abelian gauge field may be denoted as
$Z[n(x),u(x),\beta(x),\mu_i(x)]$. It is a function of velocity of the macroscopic motion $u(x)$, space - depending temperature $\beta(x)$, and varying chemical potentials $\mu_i(x)$ corresponding to the conserved charges of the system. In the present paper we imply that vector $n$ orthogonal to surface $\Sigma$ is constant $n = (1,0,0,0)$. We derived representation of this partition function in the form of Euclidean functional integral over fermionic fields, the gauge field (taken in temporal gauge), and the corresponding conjugate momentum. Besides, we represent it as an analytical continuation of partition function for the effective quantum field theory in Minkowski space - time. The latter effective theory seems to us especially instructive, and we represent its form here:   
$$
Z[n(x),u(x),\beta(x),\mu_i(x)] = {\cal Z}[n(x),u(x),\beta(x),\mu_i(x),-i],$$
where the Minkowski space partition function depends on parameter $h$. It is to be taken equal to $-i$ in order to arrive at the original statistical partition function:
\begin{equation}
	{\cal 	Z}[n,u,\beta(x),\mu_i(x),h]=\int D\bar{\psi} D\psi DA_\mu   \, e^{i\int  d^4 x \, \mathcal{L}(\overline{\psi},\psi ,A)}\label{ZZZ}
\end{equation} 
Here $x = (t,\vec{x})$. Functional integration here is over the gauge fields without gauge fixing. This allows to discretize the system and perform numerical simulations of the Wick - rotated effective theory. Alternatively one may fix any gauge using the standard Faddeev - Popov procedure, and develop the corresponding perturbation theory. Propagators of both fermions and gauge bosons in this theory differ essentially from those of the original one, and contain the frigidity vector. 
The same refers also to the interaction vertices. 

 Integral in exponent of Eq. (\ref{ZZZ}) is to be taken along the piece of space - time of extent ${\mathfrak B}(\vec{x}) h $ (in the reference frame with $n = (1,0,0,0)$) that starts from the given hyperplane $\Sigma$ corresponding to $t = t_0$. Here $\mathfrak B$ is arbitrarily chosen function of spatial coordinates. (Physical observables should not depend on this choice.)  The effective lagrangian is not relativistic invariant, it depends on the macroscopic four - velocity $u(t_0,\vec{x})$ and function ${\mathfrak B}(\vec{x})$ through the four - vector field $\mathfrak U$ given by 
\begin{equation}
	{\mathfrak U}(\vec{x}) = \frac{\beta(t_0,\vec{x})}{{\mathfrak B}(\vec{x})} u(0, \vec{x})\label{mUc}
\end{equation}
 
We have the two obvious choices of function $\mathfrak B$: 

\begin{enumerate}
	
	\item 
	${\mathfrak B}(\vec{x}) =\beta(t_0,\vec{x})$. In this case 
	\begin{equation}
		{\mathfrak U}^\mu(\vec{x}) =  u^\mu(t_0, \vec{x})\label{mU1c}
	\end{equation}
	In this case vector field $\mathfrak U$ may be interpreted as the four - velocity distribution at the initial moment, while function $\mathfrak B$ is a natural spatial depending inverse temperature. 
	
	\item
	
	 ${\mathfrak B}(\vec{x}) =\beta(t_0,\vec{x}) u^0(t_0,\vec{x})$. In this case 
	\begin{equation}
		{\mathfrak U}^\mu(\vec{x}) =  \frac{u^\mu(t_0, \vec{x})}{u^0(t_0, \vec{x})}\label{mU2c}
	\end{equation}
	Then vector field $\mathfrak U$ cannot be interpreted as four - velocity of macroscopic motion ($\mathfrak B$, though, may still be interpreted as inverse temperature).

\end{enumerate}

For the  choice of $\mathfrak U$ given by Eq. (\ref{mU2c}) the effective lagrangian is simplified:
In this case we arrive at the lagrangian 
\begin{align}
	\mathcal{L}(\overline{\psi},\psi,A)
	=& (\overline{\psi}(\gamma^\mu\frac i2\overset{\leftrightarrow}{D}_\mu-m)\psi)  
	+ \sum_i\mu_ij^{0}_i + \nonumber\\
	&{+ {\mathfrak U}_k(\overline{\psi}\gamma^0(\frac{i}{8}[\gamma^{j},\gamma^k](\overset{\leftarrow}{D}_j+D_j) 
		-\frac{i}{2} \overset{\leftrightarrow}{D^k})\psi)} + \nonumber\\
	&+ {\frac{1}{2 }\Bigl({E}^{ai}{E}^{ai}-(1- \vec{\mathfrak U}^2){B}^{ai}{B}^{ai}\Bigr)}\nonumber\\& - \frac{1}{2}({B}^{a}_j{\mathfrak U}^j)({B}^{a}_i{\mathfrak U}^i)-{\epsilon_{ijk}} {}{E}^{a}_i{B}^{a}_j{\mathfrak U}^k,\label{finL1}
\end{align}
where $E^a_i = F^a_{0i}$ is the chromoelectric field, while $B^a_i = \frac{1}{2}\epsilon_{ijk}F^{ajk}$ is the chromomagnetic field. 	
In particular, for the three most important particular cases of equilibrium macroscopic motion we have (now ${\mathfrak U}^0 = 1$, while $\vec{\mathfrak U} = \vec{u}(t_0,\vec{x})/u^0(t_0,\vec{x}) =\vec{u}(t_0,\vec{x}) \sqrt{1-\vec{u}(t_0,\vec{x})^2} $). 
\begin{enumerate}
	\item 
		Motion with constant four - velocity $u(x) = {\rm const}$. 
	\item	
	Rigid rotation with constant angular velocity $\omega$ around axis $z$:
	$$
	{\mathfrak U}(x) = \frac{1}{\sqrt{1-\omega^2(x^2 + y^2)}}(1,-y\omega,x\omega,0)
	$$
	Then 
	$$
	{\mathfrak B}(\vec{x}) = \beta_0, \quad b = (\beta_0,0,0,0)
	$$ 	
	while 
	$$
	{\mathfrak U} = (1,-y\omega,x\omega,0)
	$$
	{In Appendix \ref{comparison} we calculate in details the effective lagrangian for this case. It is then compared to the effective lagrangian obtained using the so - called passive description of rotation, in which the partition function is calculated in the rotating reference frame. It appers that both approaches result in identical lagrangians.} 
	\item 
	The initially accelerated motion with  "acceleration" $a$ along axis $x$:  
	$$
	{\mathfrak U}(t,\vec{x}) = \frac{1}{\sqrt{(1+ax)^2 - a^2 t_0^2}}(1+ax,at_0,0,0)
	$$
	For the choice $t_0=0$ we have:
	\begin{equation}
	{\mathfrak B}(\vec{x}) = \beta_0 (1+ax)
	\end{equation}
	while 
	$$
	\vec{\mathfrak U} = (1,0,0,0)
	$$
	one can see that in this case the effective lagrangian is especially simple. It is reduced to the lagrangian of the system remaining at rest. The only effect of "acceleration" is manifested through the space - dependent temperature, which becomes negative at $x < -1/a$. Appearance of negative temperature might be related somehow to Unruh effect \cite{Khakimov2023emy}.
\end{enumerate}

Average of an operator $\cal O$ depending on fields $\Psi, A$ and their spatial  derivatives can be calculated in this effective theory as
\begin{eqnarray}
\langle T {\cal O} \rangle &=&\frac{1}{\cal Z}\int D\bar{\psi} D\psi DA   \, e^{i\int  d^4 x \,  \mathcal{L}(\overline{\psi},\psi ,A)}{\cal O}\label{OOO}
\end{eqnarray}
Here $T$ means time ordering. 
In order to calculate quantum averages of observables using numerical lattice simulations in this system we should  perform direct Wick rotation, and consider the corresponding Euclidean theory. We may also come back to the representation with  statistical  partition function $Z[n(x),u(x),\beta(x),\mu_i(x)]$ (which assumes combination of Wick rotation with rescaling of time).

\subsection{Covariant form of effective action and extension of the obtained expression to the theory out of equilibrium }

Let us rewrite the effective lagrangian of Eq. (\ref{finL4}) in the following form 
{\begin{align}
		&\mathcal{L}(\overline{\psi},\psi,A)
		= ({\mathfrak U}n)(\overline{\psi}(\gamma^\mu\frac i2\overset{\leftrightarrow}{D}_\mu-m)\psi)  
		\nonumber\\&	+(1-({\mathfrak U}n))(\overline{\psi}(\gamma n)\frac{i}{2}(n\overset{\leftrightarrow}{D})\psi) + \sum_i\mu_i(j^{}_in) + \nonumber\\
		&+ {\mathfrak U}_{\bar{\rho}}(\delta_\rho^{\bar{\rho}}-n^{\bar{\rho}}n_\rho)(\delta_\sigma^{\bar{\sigma}}-n^{\bar{\sigma}}n_\sigma)(\overline{\psi}(\gamma n)\frac{i}{8}[\gamma^{{\sigma}},\gamma^\rho](\overset{\leftarrow}{D}_{\bar{\sigma}}+D_{\bar{\sigma}}) 
		\psi)\nonumber\\
		&- {\mathfrak U}_{\bar{\rho}}(\delta_\rho^{\bar{\rho}}-n^{\bar{\rho}}n_\rho)(\overline{\psi}(\gamma n)\frac{i}{2} \overset{\leftrightarrow}{D^{{\rho}}}\psi) + \nonumber\\
		&- \frac{1}{4 {\mathfrak U}n}{F}^{a\mu\nu}{F}^{a}_{\mu\nu}\nonumber\\
		&-\frac{{\mathfrak U}^2 -1}{4 {\mathfrak U}n}{F}^{a\rho\sigma}{F}^{a}_{\bar{\rho}\bar{\sigma}}(\delta_\rho^{\bar{\rho}}-n^{\bar{\rho}}n_\rho)(\delta_\sigma^{\bar{\sigma}}-n^{\bar \sigma}n_\sigma))\nonumber\\& - \frac{1}{8 {\mathfrak U}n}(n^\mu\epsilon_{\mu \nu \rho\sigma}{F}^{a\rho\sigma}{\mathfrak U}^\nu)(n^{\bar{\mu}}\epsilon_{\bar{\mu}\bar{\nu}\bar{\rho}\bar{\sigma}}{F}^{a\bar{\rho}\bar{\sigma}}{\mathfrak U}^{\bar{\nu}})\nonumber\\
		&-\frac{1}{ {\mathfrak U}n}{F}^{a}_{\nu\mu}{F}^{a\nu \rho}{\mathfrak U}_\rho n^\mu + {F}^{a}_{\nu\mu}{F}^{a\nu \rho}{n}_\rho n^\mu .\label{finL}
\end{align} }
Here we restore vector $n$ orthogonal to hyperplance $t = const$ in order to have covariant expressions. 

The derivation of Eq. (\ref{finL}) was given in the present paper for the specific types of motion admitted for thermal equilibrium. Now let us consider the case of hydrodynamic approximation, in which locally the system remains in thermal equilibrium. Therefore, locally its macroscopic motion is reduced to constant speed movement, rigid rotation, and accelerated motion. In this situation we are able to describe the given system by Eq. (\ref{finL}) with $4$ - vector of macroscopic velocity $\mathfrak U$, which may have arbitrary profile (provided that the hydrodynamic approximation remains valid). The partition function of the system is then given by Eq. (\ref{ZZZ}). Integral in exponent of Eq. (\ref{ZZZ}) is to be taken along the piece of space - time of extent ${\mathfrak B}(\vec{x}) h $ (in the reference frame with $n = (1,0,0,0)$) that starts from the given hyperplane $\Sigma$ corresponding to $t = t_0$. Here ${\mathfrak B}(\vec{x})$ is to be considered as space - dependent inverse temperature. 

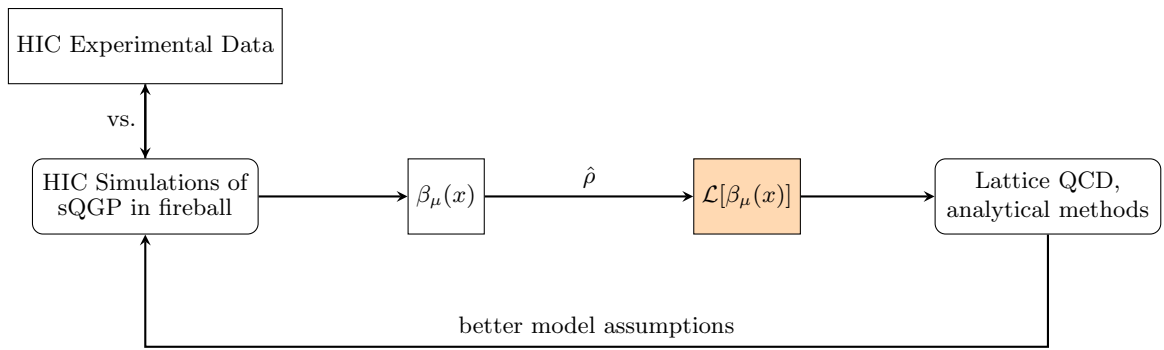
\begin{figure*}
	\begin{tikzpicture}[node distance=2cm]
		\node (PHS) [method] {\shortstack{HIC Simulations of \\ sQGP in fireball}};
		\node (BETA) [obj, right of=PHS, xshift=2cm] {\(\beta_\mu(x)\)};
		\node (LE) [obj, right of=BETA, xshift=2cm, fill=orange!30] {\(\mathcal{L}[\beta_\mu(x)]\)};
		\node (NP) [method, right of=LE, xshift=2cm] {\shortstack{Lattice QCD, \\ analytical methods}};
		\node (EXP) [obj, above of=PHS] {\shortstack{HIC Experimental Data}};
		\draw [arrow] (PHS) -- (BETA);
		\draw [arrow] (BETA) -- node[anchor=south] {\(\hat\rho\)} (LE);
		\draw [arrow] (LE) -- (NP);
		\draw [arrow] (NP) -- ++(0,-2) 
		--node[anchor=south] {better model assumptions } ++(-12,0) 
		-- (PHS);
		\draw [arrow] (PHS) -- node[anchor=east] {vs.} (EXP);
		\draw [arrow] (EXP) --  (PHS);
	\end{tikzpicture}
	\caption{Schematic block - diagram for proposed investigation of the strongly coupled quark - gluon plasma ((s)QGP) in Heavy Ion Collisions (HIC). The investigation starts from the experimental data obtained directly at the accelerators, where fireballs appear during the HIC. These data are used for the  simulations of various phenomenological models of sQGP existing within the fireball \cite{Bjorken1983,Heinz2013th,Csernai2013,Teaney2001av,Zinchenko2022tyg,Bravina2021arj,Rybalka2018uzh}. Four - vector field of frigidity $\beta_\mu(x)$ may be extracted from these simulations \cite{Bravina2021arj}. Later it is to be used to build Zubarev statistical operator    
		\(\hat\rho\) for the sufficiently small cell, in which motion is assumed to have equilibrium form (which means it is the superposition of motion with constant linear velocity, rigid rotation and motion with linear acceleration). Next, our effective lagrangian $\mathcal{L}$ of Eq. (\ref{finL}) is to be built for the obtained profile of $\beta_\mu(x)$. Various methods (both analytical and numerical)  may be used to investigate the resulting model. Results obtained during this investigation are to be used back in the HIC simulations in order to improve their assumptions, and in order to calculate various physically observed quantities.  }
	\label{FigFlowChart}
\end{figure*}

Synthesis of HIC simulations \cite{Bjorken1983,Heinz2013th,Csernai2013,Teaney2001av,Zinchenko2022tyg,Bravina2021arj,Rybalka2018uzh} and study of our effective model with the effective action of Eq. (\ref{finL}) proposes the scheme of investigations of quark - gluon plasma illustrated by FIG. \ref{FigFlowChart}.

\section{Conclusions}

In the present paper we developed the approach by Zubarev to the description of macroscopic motion of a substance described by relativistic quantum field theory with fermions and inter - fermion interactions. The latter are taken in the form usual for the relativistic quantum field theory - i.e. as the gauge theory interacting with fermions. Our conclusions, however, remain valid also for the other types of interactions (as the ones present in the description of fermion superfluid $^3$He - A with emergent relativistic invariance). In the latter case one needs, however, a certain modification of the formalism related to anisotropy given by the nontrivial vierbein defined in the inertial laboratory reference frame. 

The mentioned above types of macroscopic motion (admitted for thermodynamic equilibrium) may be considered using Zubarev statistical operator for any substance described by relativistic quantum field theory. The previously used methods of investigation relied on the operator formalism. In the present paper we develop the functional integral technique that allows to explore these types of motion. Namely, we reduce the calculation of statistical averages of various quantities with respect to the Zubarev statistical operator to the calculation of the corresponding correlation functions within the functional integral formalism, in which the integration is performed over the dynamical fermion fields and dynamical gauge fields. The effective lagrangian entering this functional integral is derived explicitly. It depends on the four -  velocity $u$ of the macroscopic motion. In the present paper we consider the simplest possible foliation of space time, in which surfaces $\Sigma_\sigma$ for any value of $\sigma$ are the hyperplanes $t = const$. It would be interesting to consider the extension of the presented formalism to arbitrary form of $\Sigma_\sigma$. 
The corresponding construction then should include Hamiltonian quantization of gauge theory in curved space \cite{Vassilevich1991rt}. Interesting observation is that physical quantities should not depend on the form of surfaces $\Sigma_\sigma$, which might result in the corresponding Ward identities. However,  this extension is out of the scope of the present paper. 

The above given consideration of rigid rotation demonstrates that our approach gives the effective lagrangian identical to the one of the system that is at rest in rotating reference frame ("passive" description of rotation) - see Appendix \ref{comparison}. Therefore, the effective action obtained in the present paper is intended to be used for the other types of motion admitted for equilibrium, which include simultaneously all three basic types of motion: motion with constant velocity, rigid rotation, and "accelerated" motion. For such a superposition of general type the "passive" description of motion (system at rest in a certain reference frame) is not known.  

We expect  that the presented path integral formulation may be used for the investigation of quark - gluon plasma, which appears in the state with local thermal equilibrium during the heavy ion collisions. The regime of relatively small chemical potential (compared to temperature) may be explored using lattice simulations. The lagrangian of Eq. (\ref{finL}) accounts for the general type of motion.  Being implemented to lattice simulation, Eq. (\ref{finL}) then might give an important information about the quark - gluon plasma in the fireballs appeared during the heavy  - ion collisions. Until now its consideration within lattice quantum field theory has been limited by the rigid rotation (see, for example, \cite{Braguta2021jgn}). In real fireballs macroscopic motion has more complicated form.  

Another possible field, where our results may be used is physics of fermionic superfluid $^3$He - A \cite{VolovikBook}. Here our methodology should be modified to include anisotropy of the system. Then  Eq. (\ref{finL}) is to be modified accordingly, and may be used for the theoretical analysis of superfluid in the presence of macroscopic motion \footnote{In view of the obvious analogy between fermionic superfluids and electronic liquids in topological semimetals \cite{burkov2016topological} we expect that the methodology developed in the present paper may also be extended even further - to the theoretical investigation of electronic properties of these materials. Especially promising is the  analysis of interplay between chiral anomaly, anomalous transport and description of macroscopic motion of electronic liquid in Weyl semimetals \cite{zyuzin2012topological}.}. This modification, however, remains out of the scope of the present paper.

\begin{appendices} 

\begin{section}{Vector field $u(x)$ for rigid rotation}
	
	\label{rigid}

In inertial laboratory reference frame (cylindrical coordinates) the 
coordinates are $(t,\phi,r,z)$ while metric is 
$$
ds^2 = dt^2 - r^2d\phi^2 - dr^2 - dz^2
$$

\begin{enumerate}

\item{Rigidly rotating reference frame. Cylindrical coordinates.}

In the rigidly rotating reference frame (around axis $z$) coordinates are $(t,\tilde{\phi},r,z)$ with $\tilde{\phi} = \phi - \omega t$. Metric is given by 
\begin{equation}
ds^2  = (1-\omega^2r^2) dt^2 - 2 r^2 \omega dt d\tilde{\phi} -r^2 d\tilde{\phi}^2 - dr^2 - dz^2 \label{rigidg}
\end{equation}

In this reference frame 
$$
u = \frac{1}{\sqrt{1-\omega^2 r^2}}(1,0,0,0)
$$

\item{Inertial reference frame. Cylindrical coordinates.}

Transition between the two reference frames is given by matrix 
$$
\Omega =
 \left(\begin{array}{cccc} 1 & 0 & 0 & 0\\
\omega & 1 & 0 & 0\\
0 & 0 & 1 & 0\\
0 & 0 & 0 & 1\end{array}	\right)
$$
Here we have 
$$
u(x) = \frac{1}{\sqrt{1-\omega^2 r^2}}(\hat{t} + \omega\hat{\phi})
$$
Vectors $\hat{t}$ and $\hat{\phi}$ are defined as unit vectors in direction of changing of variables $t$ and $\phi$.

\item{Inertial reference frame. Cartesian coordinates.}

In the Cartesian coordinates $(t,x=r {\rm cos}\,\phi,y=r {\rm sin}\,\phi,z)$ of laboratory reference frame we get (${\rm tg}\,\phi = y/x $):
$$
u(x) = \frac{1}{\sqrt{1-\omega^2(x^2 + y^2)}}(1,-y\omega,x\omega,0)
$$
One can see that this vector is not defined for $r > 1/\omega$.

Now let us suppose that the macroscopic motion of the system considered in the main text occurs with four - velocity $u(x)$. Then we have
$$
\beta_\rho(x) = \beta(x) u_{\rho}(x) = b_\rho + \bar\omega_{\rho\sigma} x^\sigma
$$
with 
$$
\beta(x) = \beta_0 \sqrt{1-\omega^2(x^2+y^2)}, \quad b = (\beta_0,0,0,0), $$$$ \bar{\omega} = \beta_0 \left(\begin{array}{cccc}0& 0& 0 & 0\\
0 & 0 & -\omega & 0\\
0 & \omega & 0 & 0\\
0 & 0 & 0 & 0 \end{array}\right)
$$
and arbitrary constant $\beta_0= 1/T_0$ related to constant $T_0$ of dimension of temperature.

\end{enumerate}

	\end{section}

\begin{section}{Vector field $u(x)$ for accelerated motion}
	
	\label{accelerate}

	\begin{enumerate}
		
		\item{Laboratory reference frame. Hyperbolic coordinates.}
		
			In inertial laboratory reference frame  the 
		coordinates are $x = (t,x,y,z)$ while metric is 
		$$
		ds^2 = dt^2 - dx^2 - dy^2 - dz^2
		$$
		
		In these coordinates the components of vector $u^k(x)$ form differential operator (i.e. vector in  tangent bundle) $u = u^k \frac{\partial}{\partial x^k}$. In the other coordinates $\tilde{x}$ the components of $u^k$ transform accordingly as $u^k \to u^l \frac{\partial \tilde{x}^k}{\partial {x}^l} $
		
		Let us introduce the new coordinates $(v, \eta, y, z)$:
		$$x = (v \, {\rm ch}\, \eta, v\, {\rm sh}\, \eta, y, z)$$
		with metric
		$$
		ds^2 = dv^2 - v^2 d\eta^2  - dy^2 - dz^2
		$$

		\item{"Accelerated" reference frame. Hyperbolic coordinates}
		
		The coordinates are $X = (v,\tilde{\eta},y,z)$ with $\tilde{\eta} = \eta - a v$. Metric is given by 
		\begin{equation}
			ds^2  = (1-a^2v^2) dv^2 - 2 v^2 a dv d\tilde{\eta} -v^2 d\tilde{\eta}^2 - dy^2 - dz^2 \label{accg}
		\end{equation}
		
		In this reference frame we define unit vector
		$$
		{u = \frac{\hat{v} {\rm ch}\, (\tilde\eta +av) + \hat{\tilde \eta}\,(a (1 - {\rm ch}\, (\tilde \eta +av)) - \frac{1}{v}\, {\rm sh}\, (\tilde \eta +av)) }{\|\hat{v} {\rm ch}\, (\tilde\eta +av) + \hat{\tilde \eta}\,(a (1 - {\rm ch}\, (\tilde \eta +av)) - \frac{1}{v}\, {\rm sh}\, (\tilde \eta +av)) \|}}$$
		It is directed along unit vector $\hat v$ in direction of changing variable $v$ at $\eta = 0$. For $\eta \ne 0$ it has also component along vector $\hat{\tilde{\eta}}$ defined as $\hat{\tilde{\eta}}= \frac{\partial}{\partial \tilde{\eta}}$. 
		
		\item{$u(x)$ in inertial reference frame. Hyperbolic coordinates.}
		
		Transition between the two reference frames is given by matrix 
		$$
		\Omega =
		\left(\begin{array}{cccc} 1 & 0 & 0 & 0\\
			a & 1 & 0 & 0\\
			0 & 0 & 1 & 0\\
			0 & 0 & 0 & 1\end{array}	\right)
		$$
		Then
		$$
		u(x) = \frac{\hat{v} \, {\rm ch}\,\eta + \hat{\eta}\, (a - \frac{{\rm sh}\, \eta}{v})}{\|\hat{v} \, {\rm ch}\,\eta + \hat{\eta}\, (a - \frac{{\rm sh}\, \eta}{v})\|}
		$$

		\item{Inertial reference frame. Cartesian coordinates.}
		
		In the Cartesian coordinates $(t=v\, {\rm ch}\,\eta,x=v\, {\rm sh}\,\eta,y,z)$ of laboratory reference frame we get (${\rm th}\,\eta = x/t $):
		$$
		u(x) = \frac{1}{\sqrt{(1+ax)^2 - a^2 t^2)}}(1+ax,at,0,0)
		$$
		One can see that all vectors of tetrad are not defined for $t > |1/a+x|$.
		
		Now let us suppose that the macroscopic motion occurs with four - velocity $u(x)$. Then we have
		$$
		\beta_\rho(x) = \beta(x) u_{\rho}(x) = b_\rho + \bar\omega_{\rho\sigma} x^\sigma
		$$
		with 
		$$
		\beta(x) = \beta_0 \sqrt{(1+ax)^2 -a^2 t^2)}, \quad b = (\beta_0,0,0,0), $$$$ \bar{\omega} = \beta_0 \left(\begin{array}{cccc}0& a& 0 & 0\\
			-a & 0 & 0 & 0\\
			0 & 0 & 0 & 0\\
			0 & 0 & 0 & 0 \end{array}\right)
		$$
		and arbitrary constant $\beta_0= 1/T_0$ related to constant $T_0$ of dimension of temperature.

	\end{enumerate}
	
\end{section}	
	
	\begin{section}{Coherent states for relativistic fermions}\label{SectAppCoherent} 

		In this section we suggest the way to introduce coherent states for Dirac fermions.
		Our aim is to build the identity resolution with a Grassmann 4-spinor fields 
		\(\psi(x),~\bar\psi(x)\) on a space-like hyperplane \(\Sigma\)
		(as in \eqref{Zub}).
		However, the Dirac field operator is not an annihilation operator for the Dirac vacuum
		\begin{gather}
			\hat\Psi(x)\ket{0}\neq 0,
		\end{gather}
		since the field operator contains the creation operators for the antiparticles
		\begin{gather}
			\hat\Psi(x) = \sum_{\lambda,\bp}\frac{1}{\sqrt{2E_p}}(
			e^{-ipx}u_{\lambda\bp}\hat a_{\lambda\bp} 
			+ e^{ipx}v_{\lambda\bp}\hat b^\dag_{\lambda\bp}),\\ 
			\sum_{\lambda,\bp}=\sum_\lambda\int\frac{d^3p}{(2\pi)^3},
			~px = E_px_0-\bp\mathbf{x},~E_p=\sqrt{\bp^2+m^2}\nonumber
		\end{gather}
		Therefore, the construction of the corresponding coherent states is complicated. 
		
			Instead, we consider a special empty state \(\Omega,~ \braket{\Omega|\Omega}=1\)
			with all the particle states vacant, and all the antiparticle states occupied.
			The empty state is in the field operator kernel,
			and the normal ordering is defined as
			\begin{gather}
				\hat\Psi(x)\ket{\Omega}=0,
				\quad :\hat\Psi_\alpha(x)\hat{\bar\Psi}_\beta(x'): = -\hat{\bar\Psi}_\beta(x')\hat\Psi_\alpha(x) \label{EqPsiNO}
			\end{gather}
			Then  we use the field operators \(\hat{\bar\Psi}^\dag_\alpha(x),~\hat\Psi_\alpha(x)\), correspondingly,
			and the Grassmann-valued fields \(\bar\psi(x), \psi(x)\).
			
			The field operators are introduced in the inertial laboratory frame, and \(\hat{\bar\Psi}=\hat\Psi^\dag\gamma_0\).
			Since points \(x,y\in\Sigma\) are separated by space-like interval or coincide \((x-y)^2\le 0\), 
			the anticommutation relations hold for the corresponding  operators
			\begin{gather}
				\{\hat\Psi_\alpha(x),\hat{\bar\Psi}_\beta(y)\}|_{x_0=y_0}
				=(\gamma^0)_{\alpha\beta}\delta^{(3)}(\mathbf{x}-\mathbf{y}).\label{EqPsiAC}
			\end{gather}
			Since \(\Sigma\) is space-like,
			the volume element component \(d\Sigma^0\) is non-zero at every point,
			while any \(d\Sigma^i\) contains a time difference,
			so the useful integral over the surface
			\begin{gather}
				\int d\Sigma_\mu(x)\gamma^\mu\delta^{(3)}(\mathbf{x}-\mathbf{y})|_{y\in\Sigma} = \gamma^0.\label{EqIntSg}
			\end{gather}
			
			Then, the coherent states may be introduced as 
			\begin{gather}
				\ket{\psi} = e^{\int d\Sigma_\mu\hat{\bar\Psi}\gamma^\mu \psi}\ket{\Omega},\quad 
				\bra{\psi} = \bra{\Omega}e^{\int d\Sigma_\mu{\bar\psi}\gamma^\mu \hat\Psi}.\label{EqPsis}
			\end{gather}
			These states are the eigenstates for the field operators
			\begin{gather}
				\hat\Psi_\alpha(x)\ket{\psi}=\psi_\alpha(x)\ket{\psi},
				\quad \bra{\psi}\hat{\bar\Psi}_\alpha(x) = \bra{\psi}{\bar\psi}_\alpha(x). \label{EqPropF}
			\end{gather}
			To prove the needed properties of coherent states we  
			 discretize space, use anti - commutation relations  \eqref{EqPsiAC},
			and apply \eqref{EqIntSg} 
			(note that in the discretized space the component \(d\Sigma_0=a^3\) by construction)
			\begin{align}
				&\hat\Psi_\alpha(x)\ket{\psi} = 
				\lim_{a\to 0} \hat\Psi_\alpha(x)\prod_{y\in\Sigma} e^{\D y_\mu\hat{\bar\Psi}\gamma^\mu\psi}\ket{\Omega}
				\nonumber\\&= \lim_{a\to 0}\prod_{y\neq x}e^{\D y_\mu\hat{\bar\Psi}_y\gamma^\mu\psi_y} 
				\hat\Psi_{x\alpha}\sum_{k=0}^4\frac{1}{k!}(\D x_\mu\hat{\bar\Psi}_{x}\gamma^\mu\psi_x)^k \ket{\Omega}\nonumber\\
				&= \lim_{a\to 0}\prod_{y\neq x}e^{\D y_\mu\hat{\bar\Psi}\gamma^\mu\psi} 
				\sum_{k=1}^4\frac{1}{k!}k\D x_0 a^{-3}\gamma^0\gamma^0\nonumber\\&\psi_\alpha(x)
				(\D x_\mu\hat{\bar\Psi}_{x}\gamma^\mu\psi_x)^{k-1}\ket{\Omega}\nonumber \\
				&= \psi_\alpha(x)\lim_{a\to 0}\prod_{y\neq x}e^{\D y_\mu\hat{\bar\Psi}\gamma^\mu\psi} 
				\sum_{k=1}^4\frac{1}{(k-1)!}(\D x_\mu\hat{\bar\Psi}_{x}\gamma^\mu\psi_x)^{k-1}\ket{\Omega}\nonumber \\
				&= \psi_\alpha(x)\lim_{a\to 0}\prod_{y\neq x}e^{\D y_\mu\hat{\bar\Psi}\gamma^\mu\psi} 
				\nonumber\\&\sum_{k=0}^4\frac{1}{k!}(\D x_\mu\hat{\bar\Psi}_{x}\gamma^\mu\psi_x)^{k} \ket{\Omega}\nonumber
				= \psi_\alpha(x)\ket{\psi}, 
			\end{align}
			where the \(k>4\) terms in the sums would be zeroes
			(field \(\psi\) has four Grassmann components \(\psi_\alpha,~\alpha=1..4\)),
			 the term added in the last line \(k=4\) is also zero.
			
			The basic properties of the coherent states read
			\begin{gather}
				\braket{\psi'|\psi} = e^{\int d\Sigma_\mu{\bar\psi'}\gamma^\mu\psi}\label{EqProp0}\\
				\hat I_\Sigma = \int\cD^{\Sigma}\bar\psi\cD^{\Sigma}\psi e^{\int d\Sigma_\mu\bar\psi\gamma^\mu\psi}\ket{\psi}\bra{\psi} \label{EqProp1},\\
				\braket{\psi'|:B(\hat{\bar\Psi}(x),\hat\Psi(x)):|\psi} = \braket{\psi'|\psi} \tilde B(\bar\psi'(x),\psi(x))\label{EqProp2},
			\end{gather}
			where the normal ordering is defined by \eqref{EqPsiNO},
			and in \(\tilde B\) the Lorentz time partial derivatives are to be replaced with the symmetrized Dirac Hamiltonian.
			
			The normalization \eqref{EqProp0} follows from the fundamental property \eqref{EqPropF}
			\begin{gather}
				\braket{\psi'|\psi} = \braket{\Omega|e^{\int d\Sigma_\mu{\bar\psi'}\gamma^\mu\hat\psi}|\psi}
				=  \braket{\Omega|e^{\int d\Sigma_\mu{\bar\psi'}\gamma^\mu\psi}|\psi}
				\nonumber\\=  \braket{\Omega|\psi}e^{\int d\Sigma_\mu{\bar\psi'}\gamma^\mu\psi}
				= e^{\int d\Sigma_\mu{\bar\psi'}\gamma^\mu\psi},
			\end{gather}
			as well as the identity resolution \eqref{EqProp1}.
			We note that the field operators commute with \(\hat I_\Sigma\)
			\begin{align}
				\hat\Psi_\alpha(x)\hat I_\Sigma &= \int\cD^{\Sigma}\bar\psi\cD^{\Sigma}\psi e^{\int d\Sigma_\mu\bar\psi\gamma^\mu\psi}\hat\Psi_\alpha(x)\ket{\psi}\bra{\psi} 
				\nonumber\\&= \int\cD^{\Sigma}\bar\psi\cD^{\Sigma}\psi e^{\int d\Sigma_\mu\bar\psi\gamma^\mu\psi}\psi_\alpha(x)\ket{\psi}\bra{\psi} \nonumber\\
				&= \int\cD^{\Sigma}\bar\psi\cD^{\Sigma}\psi 
				\left(\frac{\delta}{\delta\psi^\dag_\alpha(x)}e^{\int d\Sigma_\mu\bar\psi\gamma^\mu\psi}\right)
				\ket{\psi}\bra{\psi} 
				\nonumber\\&= \int\cD^{\Sigma}\bar\psi\cD^{\Sigma}\psi 
				e^{\int d\Sigma_\mu\bar\psi\gamma^\mu\psi}
				\ket{\psi}\left(\frac{\delta}{\delta\psi^\dag_\alpha(x)}\bra{\psi}\right) \nonumber\\
				&= \int\cD^{\Sigma}\bar\psi\cD^{\Sigma}\psi 
				e^{\int d\Sigma_\mu\bar\psi\gamma^\mu\psi}
				\ket{\psi}\bra{\psi}\hat\Psi_\alpha(x) 
				= \hat I_\Sigma\hat\Psi_\alpha(x)\nonumber
			\end{align}
			The integration measure \(\cD^{\Sigma}\bar\psi\cD^{\Sigma}\psi\) may be understood as the standard integration measure 
			over the functions of spatial coordinates. 
		
			Only the operator proportional to identity can commute with the field operators. (The constant of proportionality may be set to unity by appropriate rescaling of integration measure.) Therefore, the operator is independent of \(\Sigma,~\hat I_\Sigma \equiv \hat I\). 
			
			To prove the last property \eqref{EqProp2} we calculate the average with the definitions \eqref{EqPsis}.
			As a useful example, 
			let us calculate the average 
			\(\braket{\psi'|\int d\Sigma_\mu \beta_\nu\hat T^{\mu\nu}|\psi}\) 
			for the symmetric stress-energy tensor 
			\(\hat T^{\mu\nu}(x) = :\hat{\bar\Psi}(x)\frac i2 \gamma^{(\mu}\overset{\leftrightarrow}{\partial^{\nu)}}\hat\Psi(x):\),
			where the symmetrization convention is \(a_{(i}b_{j)} = \frac12(a_ib_j+a_jb_i)\), 
			and \(\overset{\leftrightarrow}{\partial}=\partial-\overset{\leftarrow}{\partial}\).
			The only complication is due to the time derivative,
			since the derivative obstructs the application of the equal-time anticommutators \eqref{EqPsiAC}.
			To get rid of the time derivative, 
			we apply the equations of motion in the operator form 
			and apply \eqref{EqPropF}. 
			For the standard choice of the hypersurface \(d\Sigma_\mu=d^3x\delta_{\mu 0}\),
			and equilibrium rigid rotation around the \(x_3\) axis with
			\(\beta_\mu=\delta_{\mu 0}\beta_0 -\epsilon_{0\mu\nu 3}\frac12\beta_0\Omega x^\nu\),
			(after integration by parts of one of the terms \(\sim\Sigma_{jk}\)) the expression simplifies to  
			{\begin{gather}
				\int d^3x\beta_\mu\braket{\psi'|\hat T^{0\mu}|\psi} = 
				\beta_0\int d^3x\bar\psi'\gamma^0(\gamma^j\frac i2 \overset{\leftrightarrow}{\partial_j} +m 
			\nonumber\\	+ \Omega( L_3 + \frac12\Sigma_{12}))\psi,\label{EqRotH}
			\end{gather}}
			where \(\hat L_3 = x_1(-i\partial_2)-x_2(-i\partial_1)\), 
			and \( L_3 + \frac12\Sigma_{12}= J_3\) provides the total angular momentum projection. 
			
			Another possible approach to construction of coherent states is to consider the annihilation operators \(\hat a_{\lambda\bp},\hat b_{\lambda\bp}\),
			which annihilate Dirac vacuum \(\ket{0}\), 
			with the standard anticommutators 
			\(\{\hat a_{\lambda\bp},\hat a^\dag_{\lambda'\bp'}\}=\delta_{\lambda\lambda'}\delta^3(\bp-\bp')\),
			\(\{\hat b_{\lambda\bp},\hat b^\dag_{\lambda'\bp'}\}=\delta_{\lambda\lambda'}\delta^3(\bp-\bp')\),
			\(\{\hat a^{(\dag)}_{\lambda\bp},\hat b^{(\dag)}_{\lambda'\bp'}\}=0\),
			and with the standard normal ordering \(:aa^\dag:=-a^\dag a\) with respect to \(\ket{0}\).
			Then, we would have  the projections of the Grassmann-valued 4-spinor fields 
			\(\int d\Sigma_\mu \frac{e^{ ipx}}{\sqrt{2E_p}}\bar u_{\lambda\bp}\gamma^\mu \psi\),
			\(\int d\Sigma_\mu \frac{e^{-ipx}}{\sqrt{2E_p}}\bar\psi \gamma^\mu u_{\lambda\bp}\),
			(\(u\to v\)),
			and might introduce alternative coherent states 
			which have properties similar to \eqref{EqProp0}-\eqref{EqProp2}.
	\end{section}
	
\begin{section}{Evolution in time in terms of path integral for gauge theory in temporal gauge}
	
	\label{path}
	
	Here we follow closely the analysis of non - Abelian gauge theory in temporal gauge given in \cite{ROSSI1980109}.
	We are considering the pure gauge theory, and choose the simplest foliation of Minkowski space - time, in which $\Sigma_\sigma$ is the hyperplane $t = const$. We consider the time evolution operator:
	\begin{eqnarray}
		U(t_i,t_f) & = &  T\, e^{-i \int d^4 x \, \hat{\cal H} }
	\end{eqnarray}
	Here we denote by $T$ ordering of the operators along the time axis:  
	$$
	T\, e^{-i \int d^4 x \, \hat{\cal H}} \equiv e^{-i \int_{\Sigma_{\sigma_f}} d\Sigma_\mu (d x^\mu/d\sigma) \delta \sigma \hat{\cal H}} \ldots $$$$ e^{-i \int_{\Sigma_\sigma} d\Sigma_\mu (d x^\mu/d\sigma) \delta \sigma \hat{\cal H}}\ldots e^{-i \int_{\Sigma_{\sigma_i}} d\Sigma_\mu (d x^\mu/d\sigma) \delta \sigma \hat{\cal H}} 
	$$
	where $\Sigma_{\sigma_i}$ is hypersurface $ t = t_i$ while $\Sigma_{\sigma_f}$ is hypersurface $ t = t_f$. Hamiltonian density for the gauge field is taken in the gauge $\hat{\cal A}_0 = 0$:
		$$
	\hat{\cal H} = \frac{1}{2} \, \Big(\hat{\Pi}^a_{i} \hat{\Pi}^a_{i} +   \hat{\cal B}^a_{i} \hat{\cal B}^a_{i}\Big)
	$$
	Here we rely on notations introduced in the main text in Sect. \ref{temporal}.
	Operator $\hat{\Pi}_{i}$ obeys canonical commutation relations with $\hat{\cal A}$:
	$$
	[\hat{\Pi}^a_{i}(x),\hat{\cal A}^{jb}(y)] = - i \delta^{ab}\delta^j_i \delta^{(3)}(x-y)\Big|_{x, y \in \Sigma_\sigma}
	$$
	Explicit construction of operator $\hat{\Pi}$ may be given in the representation, in which vectors of physical Hilbert space are identified with the complex - valued wave functionals $\Phi[{ A}]$. Namely, in this representation we define:
	\begin{eqnarray}
	\hat{\cal A}^{ai}(x)\Phi_\sigma[{ A}] &=& { A}^{ai}(x)\Phi_\sigma[{ A}]\nonumber\\
		\hat{\Pi}^a_{i}(x)\Phi_\sigma[{ A}] &=& -i\frac{\delta}{\delta { A}^{ai}(x)}\Phi_\sigma[{A}], \quad x \in \Sigma_\sigma
	\end{eqnarray}  
	Here $x$ belongs to the surface $\Sigma_\sigma$ with a particular $\sigma$. However, we establish isomorphism between Hilbert spaces defined for different values of $\sigma$.
	We consider the functions ${A}(x)$ for $x \in \Sigma_\sigma$ and $x^\prime \in \Sigma_{\sigma^\prime}$ as equal if they are equal as functions of the spatial coordinates $\vec{x}$. This allows to set up the isomorphism between the Hilbert spaces at different values of $\sigma$ - when the corresponding functionals are equal
	 $$
	 \Phi_\sigma[{A}^{i}(t,\vec{x})] = \Phi_{\sigma^\prime}[{A}^{ i}(t^\prime,\vec{x})]  
	 $$ 
	 Taking in mind this isomorphism, we omit below index $\sigma$ for functionals $\Phi$ and the corresponding symbols of abstract vectors.
	 
	 {It is assumed that the wave functionals are invariant under the gauge transformations dependent on spatial coordinates. This requirement is equivalent to Gauss constraint $$
	 	\hat{G}^a \Phi[A] = D_i \hat{\Pi}^{ai} \Phi[A] = 0 
 	$$ 
 One can see that operator $\hat{G}^a$ is generator of space - dependent gauge transformations.}  
	
	Let us choose the particular value of $\sigma$ - say, $\sigma = 0$. Correspondingly, we denote $\Sigma = \Sigma_0$. As in the case of ordinary quantum mechanics we define the bra and ket - vectors, and the eigenstates of momentum and coordinate:
	$$
	\braket{{ A}|{ A}^\prime} = \delta[{ A}-{ A}^\prime]
	$$
	$$
	\braket{{\Pi}|{\Pi}^\prime} = \delta[{\Pi}-{\Pi}^\prime]
	$$
	$$
	\braket{{A}|{\Pi}} = \frac{1}{\sqrt{\rm Vol}\,} e^{i \int d\Sigma { A}^{ai}(x) {\Pi}^a_i(x)}	$$
Here ${\rm Vol}\,$ is divergent constant that enters the following useful relation:
$$
\frac{1}{{\rm Vol}}\int D_{x\in \Sigma} {\Pi} \, e^{i \int d\Sigma \Pi^a_i(x) { A}^{ai}(x)} = \delta[{ A}] 
$$  

Here and above the functional delta - function is defined in such a way that for any functional $\Phi$ we have 
$$
\int D_{x\in \Sigma} {A}_1 \, \Phi[{ A}_1]\, \delta({ A}_1 - { A}_2) = \Phi[{ A}_2]
$$
Next, we define
\begin{equation}
	\braket{\Pi|\hat{\cal H}|{A}} = h(\Pi,{ A})\braket{\Pi|{A}}
\end{equation}
Direct calculation gives
$$
h(\Pi,{ A}) 
=\frac{1}{2} \, \Big({\Pi}^a_{i} {\Pi}^a_{i} +   { B}^a_{i} { B}^a_{i}\Big)
$$
We use completeness of the eigenstates of $\hat{\cal A}$ and $\hat{\Pi}$:
\begin{eqnarray}
	1 & = & \int D_\Sigma{ A} \ket{ A}\bra{ A}\label{AP}\\
	1 & = & \int D_\Sigma{\Pi} \ket{\Pi}\bra{ \Pi}\label{PP}
\end{eqnarray}
and represent
\begin{eqnarray}
&& T\, e^{-i \int d^4 x \, \hat{\cal H}} \equiv   \int D \Pi D{ A}\, \nonumber\\ && \ldots \bra{\Pi(x,\sigma)}e^{-i \int_{\Sigma_\sigma} d\Sigma_\mu (d x^\mu/d\sigma) \delta \sigma \hat{\cal H}}\ket{{ A}(x,\sigma+\delta \sigma)}\ldots\nonumber\\ & = & {\rm const}\,\int D \Pi D{ A}\, \nonumber\\ && \ldots e^{-i \int_{\Sigma_\sigma} d\Sigma_\mu (d x^\mu/d\sigma) \delta \sigma (\Pi(x,\sigma) (n \partial) { A}(x,\sigma) + h(\Pi(x,\sigma),{ A}(x,\sigma+\delta \sigma))}\ldots\nonumber
\end{eqnarray}

We come to 
\begin{eqnarray}
U(t_i,t_f) & = & \int  D A_{i}  D\Pi_{i} {\rm exp}\,\Big(i \int d^4 x \, \Big(\Pi^{ai} (n \partial) A^a_{i} \nonumber\\&& -\frac{1}{2} \, \Big(\Pi_{i}^a \Pi^a_{i} +   B^a_{i} B^a_{i}\Big)\Big) \Big) 
\end{eqnarray}
In this expression integration is over spatial components $A^a_i$ of gauge field.

	\end{section}

	\begin{section}{Gauge fixing on surface $\Sigma$. }\label{gaugefix}
		
		Here we describe the gauge fixing at the surface $\Sigma$ that completes the "temporal" gauge fixing $An=0$ described in the main text. We use the Faddeev - Popov procedure and introduce the corresponding ghost fields. 
		
		We start here from the following expression for the partition function in Minkowski space - time
		\begin{eqnarray}
			{\cal 	Z}[n,u,\beta(x),h]&=&\int D\bar{\psi} D\psi DA_i   \, e^{i\int  d^4 x \, \mathcal{L}(\overline{\psi},\psi ,A)}\label{ZZZD}
		\end{eqnarray} 
	In this expression the analogue of temporal gauge is fixed, i.e. here integration is over fields that obey $An = 0$. 	By $A_i$ we denote the components of vector potential orthogonal to $n$.
	Here ${\mathfrak B}(\vec{x})$ may be interpreted as inverse temperature.  
	Integral in exponent of Eq. (\ref{ZZZD}) is to be taken along the piece of space - time of extent ${\mathfrak B}(x) h $ that starts from the given surface $\Sigma$. Therefore, we actually deal with the temperature defined in the reference frame with $n = (1,0,0,0)$.  In the particular case of  infinitely large ${\mathfrak B}$ in Eq. (\ref{ZZZD}) the integral in exponent is along the infinite Minkowski space - time. The effective lagrangian entering  Eq. (\ref{ZZZD}) is invariant under the remnant gauge transformation defined on the surface $\Sigma$. The Faddeev - Popov procedure for fixing this remnant gauge freedom starts from insertion of unity into the functional integral for the quantum average of operator $\cal O$:
	\begin{eqnarray}
		\langle {\cal O} \rangle &=&\frac{1}{\cal Z} \int D\bar{\psi} D\psi D A_i  Dg DC \,\nonumber\\&&{\rm exp}\,\Big(i \int d^4 x \, ({\cal L}[A, \psi, \bar{\psi}] - \lambda \, {\rm Tr}\,C^2 \Big)\nonumber\\ && {\cal O}[A, \psi, \bar{\psi}]\delta({\cal F}[A^g] - C) \Delta_{FP}[A]\label{Keldysh0}
	\end{eqnarray} 
Here by $g$ we denote the gauge transformation depending on space - time coordinates.
	We chose here the three - dimensional gauge fixing condition ${\cal F}(x) = \partial_\mu A^\mu(x)$. $\Delta_{FP}[A]$ is Faddeev - Popov determinant given by 
	\begin{eqnarray}
		\Delta^{-1}_{FP}[A] = \int Dg \delta({\cal F}[A^g]-C) \label{FP0}
	\end{eqnarray} 
	The Faddeev - Popov determinant is equal to the determinant of operator with kernel 
	$$
	{\cal M}^{ab}(x,y) = \frac{\delta}{\alpha_a(x)} {\cal F}^b[A^g](y)\Big|_{g = 1}, \quad g \approx 1 + \alpha_a T^a
	$$
	We represent this determinant as an integral over ghost Grassmann - valued fields $\eta, \bar{\eta}$, and obtain:	
	\begin{eqnarray}
	&&	\langle {\cal O} \rangle = \int D\bar{\psi} D\psi D A_i D\bar{\eta} D\eta \nonumber\\&& {\rm exp}\,\Big(i \int d^4 x \, ({\cal L}[A, \psi, \bar{\psi}] - \lambda \, {\rm Tr}\, {\cal F}^2 + \bar{\eta} {\cal M} \eta \Big) {\cal O}\nonumber
	\end{eqnarray}

\end{section}

	\begin{section}{Comparison of active and passive descriptions of rotation. }\label{comparison}

Let us consider rigid rotation described in Appendix \ref{rigid}. The so - called passive description of rotation is given in rotating reference frame with 
\begin{equation}
	ds^2  = (1-\omega^2r^2) dt^2 - 2 r^2 \omega dt d\tilde{\phi} -r^2 d\tilde{\phi}^2 - dr^2 - dz^2 \label{rigidg}
\end{equation}	
Then nonzero components of metric tensor are:
\begin{eqnarray}
	&&g_{tt} = (1-\omega^2r^2), \quad g_{t\phi} = - r^2 \omega, \quad g_{\phi\phi} = - r^2\nonumber\\ && g_{rr} = -1, \quad g_{zz} = -1
\end{eqnarray}
In Cartesian coordinates:
\begin{equation}
	g_{\mu\nu} = \left(\begin{array}{cccc} 1-(x^2+y^2) \omega^2 & \omega y & - \omega x & 0\\
	\omega y & -1 & 0 & 0\\
-\omega x & 0 & -1 & 0\\
0 & 0 & 0 & -1 \end{array} \right)
\end{equation}
with ${\rm det}\, g = -1$. {This metric will be inserted used in the actions in place of the Minkowskian one.}

\subsection{Pure gauge theory}\label{SectAppPGT}

The action of pure gauge theory in this reference frame is  \cite{Braguta2021jgn}	
\begin{eqnarray}
	S^{(p)}_B &=& -\frac{1}{2 } \int d^4 x \Bigl((1-r^2 \omega^2) F^a_{xy}F^a_{xy} + (1-y^2 \omega^2) F^a_{xz} F^a_{xz}\nonumber\\ &&  + (1-x^2 \omega^2) F^a_{yz} F^a_{yz} - F^a_{yt}F^a_{yt} - F^a_{zt}F^a_{zt} - F^a_{xt}F^a_{xt}\nonumber\\&& + 2 \omega y (F^a_{xy}F^a_{yt} + F^a_{xz}F^a_{zt})\nonumber\\&& - 2 \omega x (F^a_{yx}F^a_{xt} + F^a_{yz}F^a_{zt})\nonumber\\&& - 2 \omega^2 xy F^a_{xz}F^a_{zy}\Bigr)
\end{eqnarray}	
In order to calculate the statistical sum we perform Wick rotation and perform integration over imaginary time from $0$ to $1/T$, where $T$ is constant temperature.

Description of rotation given in the main text of the present paper (based on Zubarev statistical operator) may be called active description. It results in the following effective action
\begin{align}
		&\mathcal{L}(A)
		= - \frac{1}{4 {\mathfrak U}n}{F}^{a\mu\nu}{F}^{a}_{\mu\nu}\nonumber\\
		&-\frac{{\mathfrak U}^2 -1}{4 {\mathfrak U}n}{F}^{a\rho\sigma}{F}^{a}_{\bar{\rho}\bar{\sigma}}(\delta_\rho^{\bar{\rho}}-n^{\bar{\rho}}n_\rho)(\delta_\sigma^{\bar{\sigma}}-n^{\bar \sigma}n_\sigma))\nonumber\\& - \frac{1}{8 {\mathfrak U}n}(n^\mu\epsilon_{\mu \nu \rho\sigma}{F}^{a\rho\sigma}{\mathfrak U}^\nu)(n^{\bar{\mu}}\epsilon_{\bar{\mu}\bar{\nu}\bar{\rho}\bar{\sigma}}{F}^{a\bar{\rho}\bar{\sigma}}{\mathfrak U}^{\bar{\nu}})\nonumber\\
		&-\frac{1}{ {\mathfrak U}n}{F}^{a}_{\nu\mu}{F}^{a\nu \rho}{\mathfrak U}_\rho n^\mu + {F}^{a}_{\nu\mu}{F}^{a\nu \rho}{n}_\rho n^\mu .\label{finLA}
\end{align} 
with 
$n = (1,0,0,0)$. In order to obtain constant value of imaginary time interval (in the functional expression for partition function) we choose function $\mathfrak B$ in such a way that $\mathfrak U$ is given by Eq. (\ref{mU2c}). Then ${\mathfrak U} = (1, - \omega y, \omega x , 0)$. 
We obtain the following expression for effective action: 
\begin{eqnarray}
	S^{(a)}_B &=& -\frac{1}{2 } \int d^4 x \Bigl((1-r^2 \omega^2) F^a_{xy}F^a_{xy} + (1-y^2 \omega^2) F^a_{xz} F^a_{xz}\nonumber\\ &&  + (1-x^2 \omega^2) F^a_{yz} F^a_{yz} - F^a_{yt}F^a_{yt} - F^a_{zt}F^a_{zt} - F^a_{xt}F^a_{xt}\nonumber\\&& + 2 \omega y (F^a_{xy}F^a_{yt} + F^a_{xz}F^a_{zt})\nonumber\\&& - 2 \omega x (F^a_{yx}F^a_{xt} + F^a_{yz}F^a_{zt})\nonumber\\&& - 2 \omega^2 xy F^a_{xz}F^a_{zy}\Bigr)
\end{eqnarray}
One can see that for pure gauge field both approaches give identical actions. 

\subsection{Non-interacting fermions}

Non - interacting fermions in rotating reference frame are described by the action
\begin{eqnarray}
	S_f = \int d^4 x \, \sqrt{-{\rm det}\, g_{\mu\nu}}\, \bar{\psi} \Bigl(\frac{i}{2}\,e^\mu_a \gamma^a \overset{\leftrightarrow}{\nabla}_\mu -m \Bigr)\psi
\end{eqnarray}
Here $\nabla_\mu = \partial_\mu + \Gamma_\mu$ is covariant derivative. The only nonzero component of spin connection is \cite{chernodub2017effects}
$$
\Gamma_t =  \frac{1}{4}\,[\gamma^1,\gamma^2]\,\omega
$$
The nonzero components of vierbein are
\begin{eqnarray}
	e^t_0 = e^x_1 = e^y_2 = e^z_3 = 1, \quad e^x_0 = y \omega, \quad e^y_0 = -x \omega
\end{eqnarray}
We obtain:
\begin{eqnarray}
	S_f &=& \int d^4 x \,  \Bigl[\bar{\psi} \Bigl(\frac{i}{2}\, \gamma^\mu \overset{\leftrightarrow}{\partial}_\mu -m \Bigr)\psi \nonumber\\ && 
	+\bar{\psi} \Bigl(\frac{i}{2}\, \gamma^0 (y \omega \overset{\leftrightarrow}{\partial}_x - x \omega \overset{\leftrightarrow}{\partial}_y) \Bigr)\psi \nonumber\\ && 
	+\bar{\psi} \frac{i}{4}\,\omega  \gamma^0  \, [\gamma^1,\gamma^2]\psi \Bigr]
\end{eqnarray}
This is to be compared with the effective lagrangian obtained for the active description of rotation:
\begin{align}
	\mathcal{L}(\overline{\psi},\psi)
	=& (\overline{\psi}(\gamma^\mu\frac i2\overset{\leftrightarrow}{\partial}_\mu-m)\psi)  
+ \nonumber\\
	&+ {\mathfrak U}_k(\overline{\psi}\gamma^0(\frac{i}{8}[\gamma^{j},\gamma^k](\overset{\leftarrow}{\partial}_j+\partial_j) 
		-\frac{i}{2} \overset{\leftrightarrow}{\partial^k})\psi 
\end{align}
For ${\mathfrak U} = (1, - \omega y, \omega x , 0)$ we obtain (after integration by parts):
\begin{align}
	\mathcal{L}(\overline{\psi},\psi)
	=& (\overline{\psi}(\gamma^\mu\frac i2\overset{\leftrightarrow}{\partial}_\mu-m)\psi) \nonumber\\ & 
	+\bar{\psi} \Bigl(\frac{i}{2}\, \gamma^0 (y \omega \overset{\leftrightarrow}{\partial}_x - x \omega \overset{\leftrightarrow}{\partial}_y) \Bigr)\psi  
	 \nonumber\\
	& 
	+\bar{\psi} \frac{i}{4}\,\omega  \gamma^0  \, [\gamma^1,\gamma^2]\psi 
\end{align}
One can see that the two approaches again give identical effective lagrangian

\subsection{Interacting fermions}\label{SectAppIF}

For the interacting fermions passive description of rotation gives

\begin{eqnarray}
	S_f &=& \int d^4 x \,  \Bigl[\bar{\psi} \Bigl(\frac{i}{2}\, \gamma^\mu \overset{\leftrightarrow}{D}_\mu -m \Bigr)\psi \nonumber\\ && 
	+\bar{\psi} \Bigl(\frac{i}{2}\, \gamma^0 (y \omega \overset{\leftrightarrow}{D}_x - x \omega \overset{\leftrightarrow}{D}_y) \Bigr)\psi \nonumber\\ && 
	+\bar{\psi} \frac{i}{4}\,\omega  \gamma^0  \, [\gamma^1,\gamma^2]\psi \Bigr]
\end{eqnarray}
where covariant derivative is defined as
$$
D_\mu = \partial_\mu + i A_\mu^a T^a
$$
On the other hand, the active approach to rotation through Zubarev operator results in 
\begin{align}
	\mathcal{L}(\overline{\psi},\psi ,A)
	=& (\overline{\psi}(\gamma^\mu\frac i2\overset{\leftrightarrow}{D}_\mu-m)\psi) \nonumber\\ & 
	+\bar{\psi} \Bigl(\frac{i}{2}\, \gamma^0 (y \omega \overset{\leftrightarrow}{D}_x - x \omega \overset{\leftrightarrow}{D}_y) \Bigr)\psi  
	\nonumber\\
	& 
	+\bar{\psi} \frac{i}{4}\,\omega  \gamma^0  \, [\gamma^1,\gamma^2]\psi 
\end{align}

One can see that there is no difference between the effective actions of the two approaches. Both passive and active descriptions of rotation yield identical Lagrangians. It is worth mentioning that in case of passive description the effective action contains variables (vector - potential of gauge field and fermionic spinor field) defined in rotating reference frame. At the same time the effective action of active description is written in terms of variables defined in the inertial laboratory reference frame. We cannot obtain one action from another formally applying the transformation between the two reference frames. This means that both descriptions are designed specifically in order to calculate thermodynamic quantities within statistical theory with inverse temperature $\mathcal B$ obtained after the corresponding Wick rotation. Obviously, the thermodynamic quantities (entropy, energy, pressure, etc) are identical for both approaches because the lagrangians are identical.

	\end{section}

\end{appendices}

\bibliographystyle{utphys} 
\bibliography{QFTMacroMotion.bib}

\end{document}